\input harvmac
%%%%%%%%%%%%%%
%
%Figure macros
%
\input epsf.sty
\ifx\epsfbox\UnDeFiNeD\message{(NO epsf.tex, FIGURES WILL BE IGNORED)}
\def\figin#1{\vskip2in}% blank space instead
\else\message{(FIGURES WILL BE INCLUDED)}\def\figin#1{#1}\fi
\def\ifig#1#2#3{\xdef#1{fig.~\the\figno}
\goodbreak\midinsert\figin{{#3}}%
\smallskip\centerline{\vbox{\baselineskip12pt
\advance\hsize by -1truein\noindent\footnotefont{\bf Fig.~\the\figno:} #2}}
\bigskip\endinsert\global\advance\figno by1}
\def\Fig#1{Fig.~\the\figno\xdef#1{Fig.~\the\figno}\global\advance\figno
 by1}
%%%%%%%%%%
%Journals

\def\np#1#2#3{Nucl.\ Phys.\ {{\bf #1}} {(#2)} {#3}}

\def\prl#1#2#3{Phys.\ Rev.\ Lett.\ {{\bf #1}} {(#2)} {#3}}

\def\ajou#1&#2(#3){\ \sl#1\bf#2\rm(19#3)}
\def\jou#1&#2(#3){,\ \sl#1\bf#2\rm(19#3)}
%%%%%%%%%%%%%%%%%%%%%%%%%%
%
% Other defs 
%
\font\bbbi=msbm10
\def\bbb#1{\hbox{\bbbi #1}}

\def\phat{{\hat p}}
\def\gym{g_{YM}}
\def\Gi{G^{-1}}
\def\frac#1#2{{#1 \over #2}}
\def\is{ & =}

\def\mth{M-theory}
\def\ltp{{\tilde l}_{pl}}
\def\Rele{R_{11}}
\def\pele{p_{11}}
\def\Mtp{{\tilde M}_{pl}}
\def\Xdot{{\dot X}}
\def\Adot{{\dot A}}

\def\Adotx{{\dot A}_x}
\def\Rtn{{\tilde R}_9}
\def\Mts{{\tilde M}_s}

\def\vecp{{\vec p}}
\def\vecb{{\vec b}}
\def\Mpl{M_{pl}}

\def\Xt{{\tilde X}}
\def\Xtdot{\dot{\tilde X}}
\def\Rn{R_9}

\def\lst{l_{st}}
\def\Asd{{\dot A}_\sigma}
\def\lpl{l_{pl}}
\def\ie{{\it i.e.}}
\def\hf{{1\over 2}}

\def\Xbar{{\overline{X}}}
\def\Gbar{{\overline{G}}}
\def\zbar{{\overline{z}}}
\def\wbar{{\overline{w}}}

\def\pperp{{{}_{\! \perp}}}

          % Maximum number of columns in a matrix

\def\ie{{\it i.e.}}
\def\eg{{\it e.g.}}

\def\DDD{/\kern -0..65em D}
\def\q{/\kern -0.65em \tilde A}
\def\qi{/\kern -0.65em \bar D}
\def\ppp{I\kern -0.65em O}
\def\calA{{\cal A}}
\def\roughly#1{\raise.3ex\hbox{$#1$\kern-.75em\lower1ex\hbox{$\sim$}}}
%%%%%%%%%
%
%References
%
\lref\Motl{L. Motl, ``Proposals on nonperturbative superstring
interactions,'' hep-th/9701025.}
\lref\Polc{J. Polchinski, ``Dirichlet-branes and Ramond-Ramond charges,'' 
hep-th/9510017\jou Phys. Rev. Lett. &75 (95) 4724.}
\lref\Tayll{W. Taylor, ``Lectures on D-branes, gauge theory and
M(atrices),'' hep-th/9801182.}
\lref\Bank{T. Banks, ``Matrix theory,'' hep-th/9710231.}
\lref\BiSu{D. Bigatti and L. Susskind, ``Review of matrix theory,''
hep-th/9712072.} 
\lref\Witt{E. Witten, ``Bound States Of Strings And $p$-Branes,'' 
hep-th/9510135\jou Nucl. Phys. &B460 (96) 335.}
\lref\PoPo{J. Polchinski and P. Pouliot, ``Membrane 
scattering with M-momentum transfer,''  
hep-th/9704029\jou Phys. Rev. &D56 (97) 6601.}
\lref\ArFr{G. Arutyunov and S. Frolov, ``Virasoro amplitude from the
$S^N\ R^{24}$ orbifold sigma model" hep-th/9708129.}
\lref\BeCo{D. Berenstein and R. Corrado, ``M(atrix)-theory in various
dimensions,'' hep-th/9702108\jou Phys. Lett. &B406 (97) 37.}
\lref\Wynt{T. Wynter, ``Gauge fields and interactions in matrix string
theory,'' hep-th/9709029\ajou Phys. Lett. &B415 (97) 349.}
\lref\DaMa{S. R. Das and S. D. Mathur, ``Interactions involving
D-branes,'' hep-th/9607149\jou Nucl. Phys. &B482 (96) 153}
\lref\HaKl{A. Hashimoto and I R. Klebanov, ``Decay of excited D-branes,''
hep-th/9604065\jou Phys. Lett. &B381 (96) 437.}
\lref\BGL{V. Balasubramanian, R. Gopakumar and F. Larsen, ``Gauge theory, 
geometry and the large N limit,'' hep-th/9712077.}
\lref\Shen{S. H. Shenker, ``Another Length Scale in String Theory?''
hep-th/9509132.}
\lref\DFS{U. H. Danielsson, G. Ferretti, and B.  Sundborg ``D-particle
dynamics and bound states,'' hep-th/9603081\jou Int. J. Mod. Phys. &A11 
(96) 5463.}
\lref\KaPo{D. Kabat and P. Pouliot, ``A comment on zero-brane quantum
mechanics,'' hep-th/9603127\jou Phys. Rev. Lett. &77 (96) 1004.}
\lref\DVV{R. Dijkgraaf, E. Verlinde, and H. Verlinde, ``Matrix string
theory,'' hep-th/9703030\jou Nucl. Phys. &B500 (97) 43.}
\lref\BaSe{T. Banks and N. Seiberg, ``Branes from matrices,''
hep-th/9612157\jou Nucl. Phys. &B490 (97) 91.}
\lref\Tayl{W. Taylor, ``D-brane field theory on compact spaces,''
hep-th/9611042\jou Phys. Lett. &B394 (97) 283.}
\lref\Seib{N. Seiberg, ``Why is the Matrix model correct?''
hep-th/9710009\jou Phys. Rev. Lett. &79 (97) 3577.}
\lref\Sen{A. Sen, ``D0 branes on $T^n$ and matrix theory,''
hep-th/9709220.}
\lref\SBGTasi{S.B. Giddings, ``Fundamental strings,''
in {\sl Particles, Strings, and Supernovae},
proceedings of the 1988 Theoretical Advanced Study Institute, 
Brown University, eds.~A.~Jevicki and C.-I.~Tan, World Scientific (1989).}
\lref\GiWo{S.B. Giddings and S. Wolpert, ``A triangulation of moduli space from 
light-cone string theory,''\ajou Comm. Math. Phys.  &109 (87) 177.}
\lref\AiSe{P.C. Aichelburg and R.U. Sexl, ``On the gravitational field of a
massless particle,''\ajou Gen. Rel. Grav. &2 (71) 303.}
\lref\BBPT{K. Becker, M. Becker, J. Polchinski, and A. Tseytlin, ``Higher 
order graviton scattering in M(atrix) theory,'' 
hep-th/9706072\jou Phys. Rev. &D56 (97) 3174.}
\lref\BeBe{K. Becker and M. Becker, ``A two-loop test of M(atrix) theory,''
hep-th/9705091\jou Nucl. Phys. &B506 (97) 48.}
\lref\BaSu{T.Banks and L.Susskind, ``Brane - anti-brane forces,''
hep-th/9511194.}
\lref\GrVh{M.B. Green and P. Vanhove, ``D-instantons, Strings and
M-theory,'' hep-th/9704145\jou Phys. Lett. &B408 (97) 122.}
\lref\GGV{ M.B. Green, M. Gutperle, and P. Vanhove, ``One loop in eleven
dimensions,'' hep-th/9706175\jou Phys. Lett. &B409 (97) 177.}
\lref\KVK{E. Keski-Vakkuri and P. Kraus, 
``M-momentum transfer between gravitons, membranes, and fivebranes as
perturbative gauge theory processes,'' hep-th/9804067.}
\lref\ClHa{M. Claudson and M. B. Halpern, ``Supersymmetric ground state
wave functions,''\ajou Nucl. Phys. &B250 (85) 689}
\lref\Flum{R. Flume, ``On quantum mechanics with extended supersymmetry and nonabelian
gauge constraints,''\ajou Ann. Phys. &164 (85) 189.}
\lref\BRR{M. Baake, M. Reinicke, and V. Rittenberg, ``Fierz identities for
real
Clifford algebras and the number of
supercharges,''\ajou J. Math. Phys. &26 (85) 1070.}
\lref\witt{E.~Witten, \np {\bf B443}(1995) 85-126, hep-th/9503124.}
\lref\BFSS{T.~Banks, W.~Fischler, S.H.~Shenker, and L.~Susskind, 
``M Theory as a matrix model: a conjecture,'' hep-th/9610043\jou
Phys. Rev. &D55 (96) 5112.}
\lref\pol{J.~Polchinski, \prl {\bf 75}(1995) 4724-4725, hep-th/9510017.}
\lref\wittd{E.~Witten, \np {\bf B460}(1996) 335-350, hep-th/9510135.}
\lref\DKPS{M.R.~Douglas,
D.~Kabat, P.~Pouliot and S.H.~Shenker, ``D-branes and short distances in
string theory,''hep-th/9608024\jou Nucl. Phys. &B485 (97) 85.}
\lref\mrd{M.R.~Douglas, hep-th/9707228.}
\lref\thooft{G.~'t Hooft, \np {\bf B190}(1981) 455-478.}
\lref\mand{S.~Mandelstam, Prog. of Theor. Phys. Suppl. No. 86 (1986)
163-170.}
\lref\gsw{M.B.~Green, J.H.~Schwarz, E.~Witten, Superstring Theory,
Cambridge University Press (1987).}
\lref\kni{V.G.~Knizhnik, Sov. Phys. Usp. {\bf 32}(11)(1989) 945-971.}
\lref\GrMe{D.~Gross and P.~Mende\jou Phys. Lett. & 197B (87) 129;
\ajou Nucl. Phys. &B303 (88) 407.}
\lref\Ven{D. Amati, M. Ciafaloni, and G. Veneziano, Phys. Lett. {\bf B197} 
(1987) 81.}
\lref\Greo{M. B. Green, ``Effects of D-instantons,'' 
hep-th/9701093\jou Nucl. Phys. &B498 (97) 195.}
\lref\Gret{M. B. Green, ``Connections between M-theory and superstrings,'' 
hep-th/9712195.}
\noblackbox
\Title{\vbox{\baselineskip-12pt\hbox{UCSBTH-98-3}\hbox{hep-th/9804121}
%\hbox{Draft -- do not distribute}
}}
{\vbox{\centerline {High Energy Scattering and D-Pair Creation}
\medskip
\centerline{in Matrix String Theory}}}
\centerline{{Steven B. Giddings}\footnote{$^*$}{giddings@physics.ucsb.edu, 
hacquebo@phys.uva.nl, verlinde@phys.uva.nl}}
\centerline{\sl Department of Physics}
\centerline{\sl University of California}
\centerline{\sl Santa Barbara, CA 93106-9530}
\medskip
%\medskip
\centerline{{Feike Hacquebord$^*$ %\footnote{$^\dagger$}{hacquebo@phys.uva.nl} 
and Herman Verlinde$^*$}}%\footnote{$^\sharp$}{verlinde@phys.uva.nl}}}
\centerline{\sl Institute for Theoretical Physics}
\centerline{\sl University of Amsterdam}
\centerline{\sl Amsterdam, Netherlands} 
\vskip.1in
\bigskip
\centerline{\bf Abstract}
In this paper we use the matrix string approach to begin a study of high
energy scattering processes in M-theory.  In particular we exhibit an
instanton-type configuration in 1+1 super-Yang-Mills theory that can be
interpreted as a non-perturbative description of a string interaction. This
solution is used to describe high energy processes with non-zero
longitudinal momentum exchange, in which an arbitrary number of eigenvalues
get transferred between the two scattering states. We describe a direct
correspondence between these semi-classical SYM configurations and the
Gross-Mende saddle points. We also study in detail the pair production of
D-particles via a one-loop calculation which in the 1+1D gauge theory
language is described by the (perturbative) transition between states with
different electric flux.  Finally, we discuss a possible connection between
these calculations in which D-particle production gives important
corrections to the Gross-Mende process.

\Date{}
%\draft
%\eqn\sym{
%S = \frac{1}{ 2\pi } \int \!d\tau d\sigma\, \tr\left((D_\a X^I)^2 + 
%\theta^T \gamma^\a D_\a \theta +
%F_{\a\b}^2 - [X^I,X^J]^2 +\theta^T\gamma_I [X^I,\theta]\right).}

%\baselineskip=13pt plus 2pt minus 1pt

\newsec{Introduction and Summary of Results}

High energy processes in string theory were first considered 
from the point of view of conventional string perturbation theory 
by Gross and Mende \refs{\GrMe}\ in the regime of fixed angle scattering
and in the near forward regime by Amati et al in \refs{\Ven}.   
The recent insights from M-theory, however, have
provided a large number of new non-perturbative tools which can
now be used to put these works into a new perspective, and extend
the results into new directions. For instance,
it was long believed that the string length $\ell_s$ marks the minimal 
distance that can be probed via scattering processes in string theory. 
This belief was based on the fact that fundamental strings tend to 
increase in size when boosted to high energies, and thus appear to be
incapable of penetrating substringy distance scales. Since the discovery 
of D-particles as non-perturbative solitons of the IIA theory, however, 
we know that there exists small scale structure that, at least for weak 
string coupling, extends well below the string length
\refs{\Shen\DFS\KaPo-\DKPS}. 
This particular realization provided important motivation for
the Matrix theory conjecture of \refs{\BFSS}\foot{For  reviews see 
\refs{\Bank\BiSu-\Tayll}
and references therein.}
that all localized excitations 
of M-theory  (including the fundamental strings) are representable as 
multi-D-particle bound states\refs{\Polc,\Witt}.

In this paper we begin a study of high energy processes in type IIA 
string theory, by making use of this Matrix theory formalism.
We focus on the four graviton scattering amplitude, and in particular 
we will present a detailed calculation of the pair production rate of 
D-particles via this process. Our aim is to probe in this way the 
transition region between the conventional perturbative string regime 
and the strong coupling regime described by 11-dimensional M-theory. 
(See fig. 1.) 

{}From the ten dimensional perspective of IIA string 
theory, D-pair production is an inelastic scattering process, in which 
two strings exchange one unit of D-particle charge. It is inherently 
nonperturbative and thus inaccessible to conventional perturbative methods.
It is also inaccessible in the traditional Matrix theory approach since
the anti-D particles are boosted to infinite energy.
{}From the eleven dimensional perspective, on the other hand, 
the D-pair creation process can simply be thought 
of as the elastic scattering of two particles in which one unit of 
Kaluza-Klein momentum in the 11 direction is exchanged. Via this 
interpretation, one can rather straightforwardly obtain a tree level 
estimate of the probability amplitude. This estimate should be reliable
for large values for the $S^1$ compactification radius $R_{11}$ and for 
collision energies sufficiently below the 11-dimensional Planck energy. 
At high energies and/or small values for $R_{11}$, on the other hand,
we expect the physics of the scattering process to be quite different 
from (semi-)classical supergravity. In the following we will attempt 
to gain more insight into this regime via the Matrix string approach.

\ifig{\fig}{The phase diagram of the $S^1$ compactification of M-theory,
with horizontal axis the log of the length scale and vertical axis the 
log of the string coupling. The various perturbative and low energy limits are 
indicated. The shaded region marks the regime where D-pair production 
is expected to be the dominant high energy process.   
}{\epsfysize=7.5cm\epsfbox{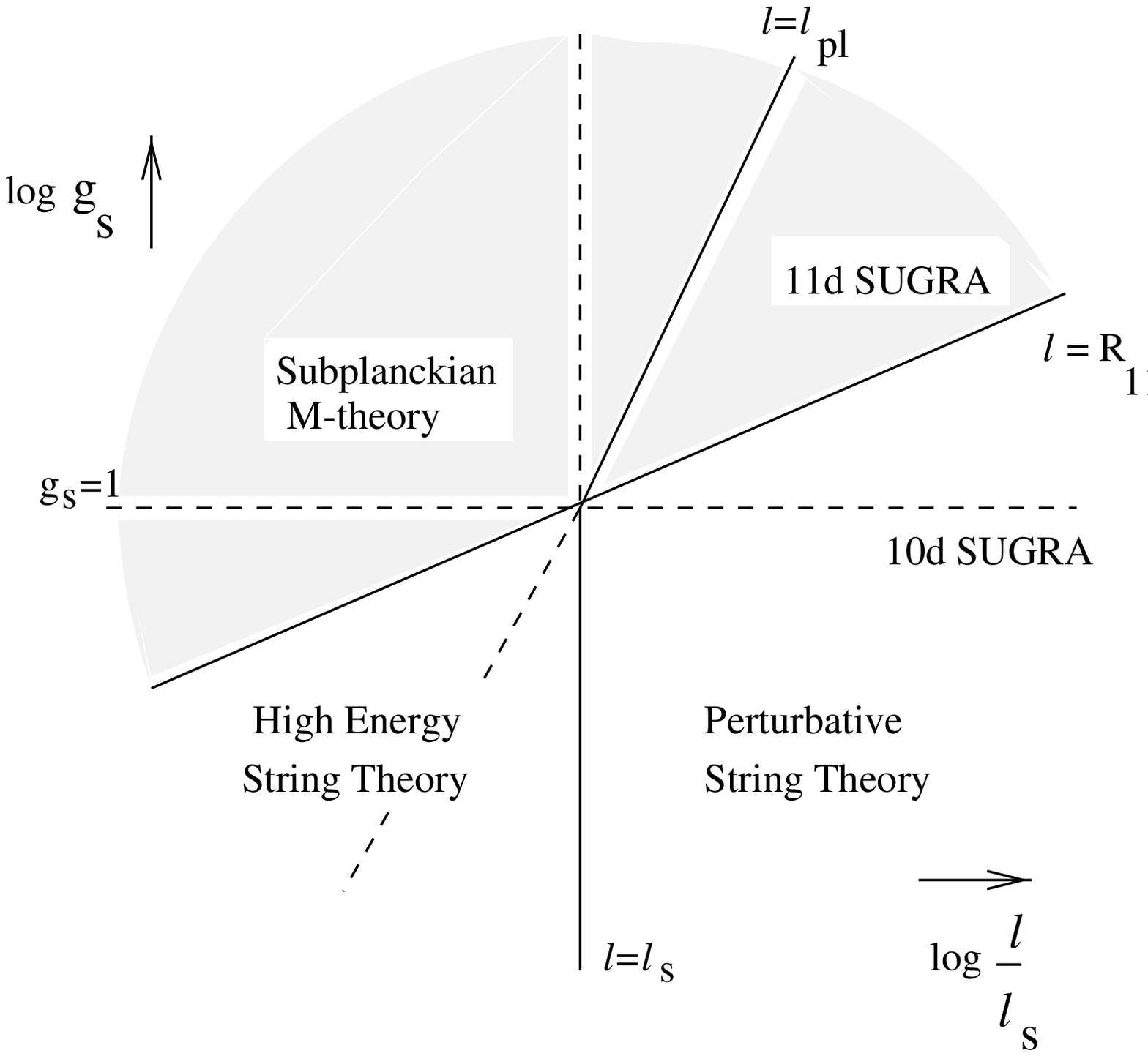}}

Matrix string theory arises from the original Matrix theory proposal 
\refs{\BFSS} via compactification on a circle, and starts from the action 
of 1+1-dimensional maximally supersymmetric Yang-Mills theory with gauge group 
$U(N)$,
\eqn\finact{\eqalign{S= \int \! d\tau\! \int_0^{1} \! \! {d\sigma} \, 
{\rm Tr} \Bigl\{- 
{g_s^2\over 4} F_{\mu\nu}^2 - \hf (D_\mu X)^2 + {1\over 4g_s^2}[X,X]^2\qquad
\cr
\qquad \quad + i{\bar \psi}\not\kern-0.3em D\psi - {1\over g_s}
\bar\psi\Gamma^i\left[X_i, \psi\right]\Bigr\}\ , }}
defined on the circle $0\leq \sigma < 1$.  The derivation of this action
is reviewed in the Appendix; here and henceforth we work in string units,
$l_s=1$.  Via the identification of the eigenvalues of the matrices $X^I$
with the transverse location of type IIA supersymmetric strings, this SYM
model can be reinterpreted as a non-perturbative formulation of light-cone
gauge IIA string \refs{\Motl\BaSe-\DVV}.  In this correspondence, the
string coupling constant $g_s$ is inversely proportional to the Yang-Mills
coupling $g_{{}_{\! YM}}$ and the free string limit therefore arises in the
strong coupling limit of the Yang-Mills model.  This correspondence has been
worked out in some detail in \refs{\DVV}.  More generally, however, all
regimes of the $S^1$ compactification of
M-theory, as indicated in fig 1, should according to the Matrix string
conjecture of \refs{\BFSS,\Motl,\BaSe,\DVV} via the above identifications
be described by particular regimes of (the large $N$ limit) of the 1+1D 
supersymmetric gauge theory.

In particular, it is expected that in the weak coupling, moderate energy
limit of the SYM theory it effectively reduces to the Matrix quantum
mechanics description of 11-dimensional supergravity. Indeed, a new feature
of matrix string theory (relative to standard light-cone string theory) is
that via the electric flux of the gauge field, the string states can be
adorned with an extra quantum number, identified with the D-particle charge
\refs{\DVV}.  In a small $g_s$ expansion, these flux sectors energetically 
decouple, corresponding to the fact that
D-particles can not be produced via perturbative string interactions.
Nonetheless, electric flux can get created in the gauge theory:
it is easy to see that electric flux creation is a simple one-loop
effect that takes place whenever a virtual pair of charged particles gets
created and annihilated, after forming a loop that winds one or more times 
around the $\sigma$ cylinder.

In the following we will develop a new method for studying high energy
scattering and D-pair production in Matrix string theory, which
will be based on a semi-classical expansion from the SYM perspective.
An important novelty of this method is that it applies to processes with 
arbitrary longitudinal momentum exchange. 
In the gauge theory language, this means that 
the transitions between the initial 
and final states that we will consider will involve a non-perturbative 
tunneling process in which an arbitrary number 
of eigenvalues get transferred between the two scattering states. 
Most previous calculations in Matrix theory relied on perturbative
SYM corrections and thus were necessarily restricted to zero $p^+$ 
transfer\foot{In \refs{\PoPo} Polchinski and Pouliot analyzed graviton
scattering with non-zero M-momentum transfer in Matrix theory.  
In their case,  the M-momentum was identified with the magnetic 
flux of the SYM gauge theory, and the corresponding instanton was 
a magnetic monopole. Here we will consider different kind of  momentum
transfer, namely of longitudinal momentum represented 
by the size  $N$ of the Matrix bound states, {\it i.e.} the number of 
D-particles in the original Matrix dictionary of \refs{\BFSS}. 
This will require a different, less familiar type of instanton
process.}.

Concretely, we will construct SYM saddle point configurations that
will allow us to interpolate between ingoing matrix configurations 
of the form
\eqn\instate{{\vec X}_{in}(\tau) = \hf
\left(\matrix
{ ({{\vec p_1}\over N_1}\tau + {\vec b})I_1 &0\cr
0&({{\vec p_2}\over N_2}\tau -{\vec b})I_2\cr}\right)\ }
and outgoing configurations of the form
\eqn\outstate{{\vec X}_{out}(\tau) =
\left(\matrix
{ ({{\vec p_3}\over N_3}\tau + {\vec b})I_3 &0\cr
0&({{\vec p_4}\over N_4}\tau -{\vec b})I_4 \cr}\right)\ }
where $I_i$ are $N_i {\times} N_i$ identity matrices, where all $N_i$'s
are {\it different} (but subject to the momentum constraint 
$N_1\! +\! N_2 = N_3\! +\! N_4$).  
These {\it in} and {\it out} configurations each
describe two widely separated gravitons with different light-cone 
momenta $p^+_{(i)} = N_{(i)}/R$ and transverse momenta $\vecp_{(i)}$, 
and with relative impact parameter ${\vec b}$. 

The interpolating solutions that we will construct, essentially look
like an appropriate matrix generalization of perturbative string
worldsheets.  The importance of these solutions is not entirely obvious,
however, since a priori one would expect that the range of validity of the
semi-classical Yang-Mills approximation has no overlap with that of
perturbative string theory. Indeed, as emphasized in \refs{\DVV} the two
regimes appear related via a strong/weak coupling duality.  However, as we
will argue in the following, even at small or moderate string coupling
$g_s$, at sufficiently high collision energies and/or impact parameters one
enters a regime in which the semi-classical SYM methods may provide an
accurate description of the scattering process.

Just like string/M-theory, the 1+1 SYM model contains various length
scales: (i) the circumference of the cylinder (set equal to $1$),
(ii) the scale set by the Yang-Mills coupling $\ell_{{}_{YM}} = 1/g_{{}_{YM}}$
\eqn\ymcoup{
\ell_{{}_{YM}}\simeq g_s}
(iii) 
the typical mass scale set by the Higgs expectation values of the SYM model. 
The latter length scale is inversely proportional to the
impact parameter $b$ of the string/M-theory scattering
process:
\eqn\bscale{
\ell_{b} \simeq {g_s\over b}\ .}
%  
%There is also another scale arising from the instanton analysis that we
%will perform later in the paper. 
Finally, (iv) there is also the length scale $\ell_E$ 
determined by the typical size of the SYM energy $E$, 
which is related to the relative space-time momenta via $E\simeq p^2/N$. 
%\eqn\Escale{
%\ell_{E} \simeq {N\over p^2}.}

The existence of these scales allows us to find small
dimensionless ratios that may parameterize the strength of the SYM
processes taking place at that scale.  For example, 
while
$g_{{}_{YM}}=1/g_s$ defines the effective coupling of SYM processes 
that take place at the scale of the YM cylinder, we also have
\eqn\gYMb{
g^{eff}_{{}_{\! YM}}(b)\, \simeq \,
 \ell_b/\ell_{{}_{YM}} \, \simeq \, 1/ b,}
as the dimensionless coupling at the scale $\ell_b$.  
%
%Corresponding to these various length scales, there are an equal number
%of dimensionless ratios that parameterize the effective strength of the
%SYM interactions at these length scales. Specifically, while
%$g_{{}_{YM}}=1/g_s$ defines the effective coupling of SYM processes 
%that take place at the scale of the YM cylinder, we also have
%%
%\eqn\gYMb{
%g^{eff}_{{}_{\! YM}}(b)\, \simeq \,
% \ell_b/\ell_{{}_{YM}} \, \simeq \, 1/b,}
%%
%and
%%
Similarly, we can also associate an effective coupling
$g^{eff}_{{}_{\! YM}}(E)$ with the scale set by the
SYM energy $E$.
%\eqn\gYME{
% \, \simeq \,
%\ell_E/\ell_{{}_{YM}}\, \simeq \, {N\over g_s p^2}. 
%}
%%
This suggests the possibility that even if $g_s$ is small or of order 1,
processes at these other 2D length scales can be accurately described by
perturbative and/or semi-classical SYM methods. This will require however
that we consider the limit of high collision energies and sufficiently
large impact parameters.\foot{In this context it may be of relevance that 
in classical 10-dimensional DLCQ supergravity, the impact parameter 
$b$ scales with the transverse relative momentum $p$ via $b 
\simeq \left(g_s^2 p^2/ N \sin \theta \right)^{1/6}$ with $N$ the DLCQ 
$p_+$-momentum.
Hence, at least in this classical context, and for fixed scattering angles 
$\theta$ and $g_s$ of order 1, the condition that $b$ is large is 
automatically satisfied in limit of large $p^2\gg N$.}

%%%%%%%%%%%%%%%%%%%%%%%%%%%%%%%%%%%%%%%%%%%%%%%%%%%%%%%%%%%%%%%%%

\ifig{\fig}{The duality chain that illustrates the dual interpretations 
of the rank N and electric flux E in matrix string theory.}
{\epsfysize=4.5cm\epsfbox{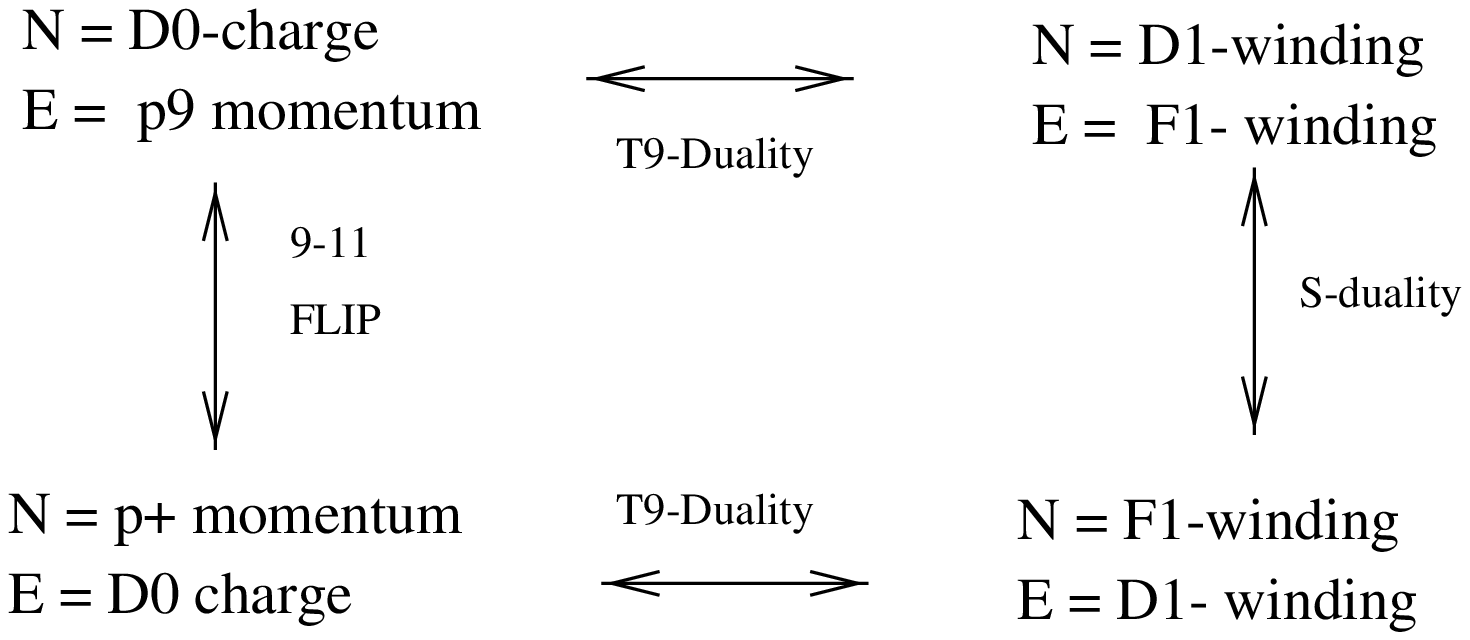}}

%%%%%%%%%%%%%%%%%%%%%%%%%%%%%%%%%%%%%%%%%%%%%%%%%%%%%%%%%%%%%%%%%

The two types of processes that we will consider, high
energy scattering with non-zero $\Delta p_+$ and the 
D-pair production, may at first sight seem quite unrelated.
% The first is a Matrix string generalization of
%Gross and Mende's treatment of fixed angle scattering.  
%This generalization arises from an instanton in the two-dimensional 
%SYM theory.  The second is the pair creation of
%D-charge.  From the string point of view this is an intrinsically
%non-perturbative process.  In the SYM theory on the other hand it is
%described by a perturbative process
%in which one of the massive charged states winds around the cylinder,
%thereby creating an electric flux.  
However, there are several connections between these two types of 
processes. First of all, it is worth pointing out that in both 
cases the scattering process involves (depending on which duality frame
one chooses) the transfer of D-particle charge and/or momentum between 
the two scattering particles. Indeed, the rank $N$ started out as 
identified with D-particle charge, and only after the duality 
it and the electric flux $E$ are mapped onto each other under an 11-9 flip: 
i.e. the interchange of the 11-th and 9-th direction. (See fig 2.)
Hence quantitative understanding both types of processes will
have a direct bearing on the Lorentz invariance of the Matrix
formalism.

Another connection between the two calculations is related to
a fundamental puzzle in the original calculation of \GrMe, namely 
the apparently dominant contribution of arbitrarily high genus 
to scattering amplitudes.  The saddle point trajectory at loop 
order $G$ typically describes a process as depicted in fig. 3: two 
incoming strings, that are wound $N=G+1$ times, interact and then 
propagate as $N$ intermediate short strings. The $N$ strings then  
join together again, producing a final state of two different 
$N$ times wound outgoing strings (see fig. 3). It was found in 
\refs{\GrMe} that the contributions of these higher order interactions 
grows larger with the genus $G$. This instability appears to signal a 
fundamental breakdown of conventional string  perturbation theory in the 
high energy regime.

%%%%%%%%%%%%%%%%%%%%%%%%%%%%%%%%%%%%%%%%%%%%%%%%%%%%%%%%%%%%%%%%%

\ifig{\fig}{This figure depicts a typical saddle point trajectory
that contributes to the high energy scattering amplitude of
fundamental strings, according to the perturbative physical picture 
proposed in \refs{\GrMe}.}
{\epsfysize=3.2cm\epsfbox{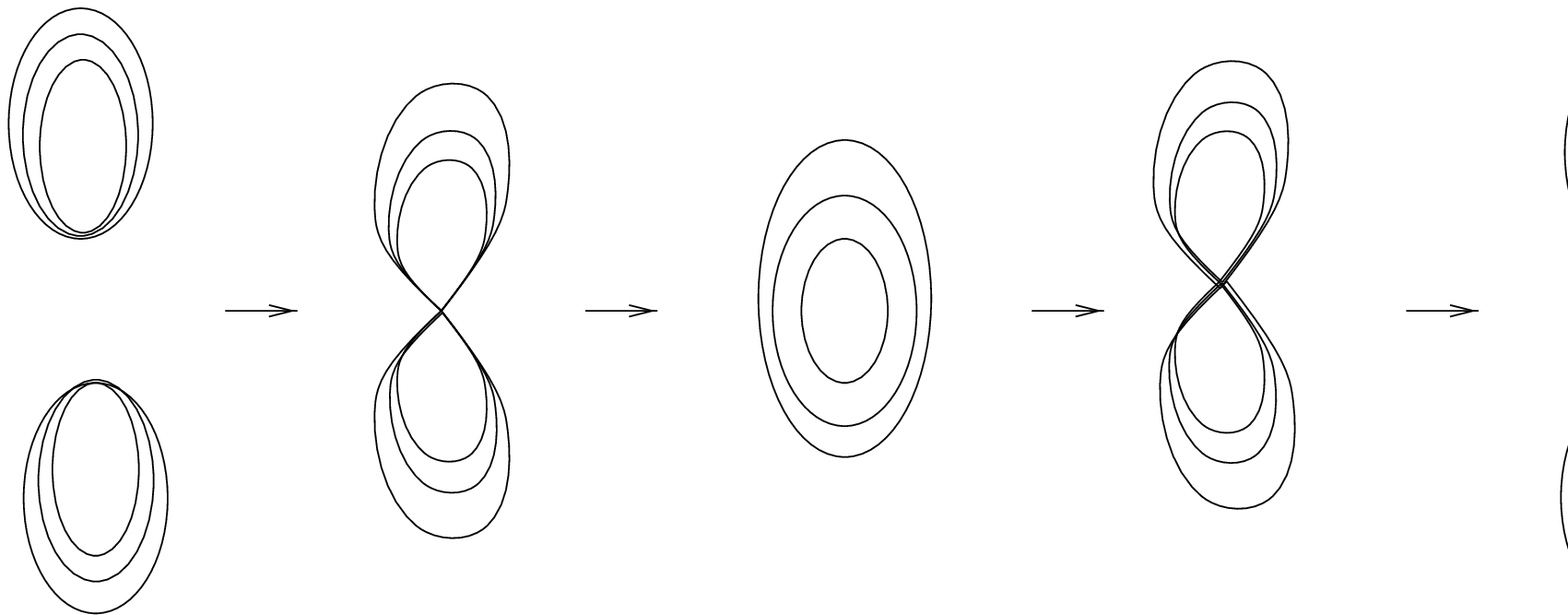}}

%%%%%%%%%%%%%%%%%%%%%%%%%%%%%%%%

On the other hand, the fragmented form of the intermediate state 
in fig. 3 gives a strong hint of some underlying non-perturbative 
structure that looks quite similar to that of the multi-D-particle 
bound state dynamics of Matrix theory.  This suggests that the
Matrix treatment may provide a rather natural stabilizing mechanism 
for a cutoff on the genus. Furthermore, our study will show that
D-particle pair production becomes relevant at this cutoff -- when 
the strings become maximally fragmented. This leads us to
suspect a deeper relation between Gross and Mende's high energy, fixed
angle scattering and the non-perturbative process of D-pair creation.

We'll begin this paper with a quick review of some properties of Matrix
strings.  We then turn to fixed angle scattering.  Section three will
recall aspects of fixed angle scattering in the traditional string
framework, with particular emphasis on its description in light-cone
gauge.  This will be followed by a discussion of string interactions in the
Matrix string formulation.  In particular, we will find a local 
instanton solution in the two-dimensional theory that describes the 
splitting/joining interaction.
Furthermore, the condition for this instanton to be matched to the
incoming/outgoing states is precisely that we be working at the world-sheet
moduli corresponding to the saddlepoint surfaces of \GrMe.  

We then turn to D-pair production, which we consider both from the
supergravity perspective as well as via a one loop
Yang Mills calculation valid for arbitrary $N$.  This is followed
by a discussion of the ranges of validity of our calculations.  In
particular, we suggest that the ranges of validity of the two calculations
overlap, and allow the picture we've mentioned in which they complement
each other.  This connection is further discussed in the final section,
together with some other observations and speculations.  An appendix
contains discussion of the derivation of the Matrix string formalism
following the approach of \refs{\Sen,\Seib}.

\newsec{Matrix String Theory}

In this section we recall some basic features of the matrix string
approach.  This approach arises from DLCQ matrix theory quantized
on a circle of ``radius'' $R$ by compactifying on another circle, and
interpreting this additional compactification as the route by which
M-theory descends to type IIA string theory.  For example, D0 branes
are states with non-zero momentum in this extra direction.

\subsec{Summary of Matrix Strings}

The Matrix string lagrangian is given in eq.~\finact.  Its form can 
be derived from M-theory on $S^1$, by combining Sen and Seiberg's 
arguments\refs{\Sen, \Seib} with the compactification prescription 
of \refs{\Tayl}. (These arguments, and a more complete discussion of 
the relation among the parameters of the theory, are presented in the Appendix.)
The basic dynamical variables of the theory are $N\times N$ hermitian
matrices and include the 8 scalar fields $X^I$ and the 8 fermion fields
$\psi^a_L$ and $\psi^{\dot a}_R$.  The Yang-Mills time variable $\tau$ 
is related to target space time by $\tau=tR$ in units where $\lst=1$.  

The $\sigma$ direction corresponds to the T-dual of the compact direction 
(herein called $x^9$) of M-theory. The target-space coordinate in this circle 
direction is identified with the zero mode of the gauge field $A_\sigma$. The 
gauge equivalence under quantized shifts
\eqn\asigma{A_\sigma \equiv A_\sigma +2\pi m} 
with $m\in {\bbb Z}$ translates via the identification of $A_\sigma$ with
the compact $X^9$ (see Appendix) 
% 
%\eqn\tilder{\widetilde{R} = \frac{1}{R_{11}} }
%
into the periodic boundary condition along the M-theory circle.

We can turn off the string interactions by considering only SYM 
configurations that describe widely separated strings, {\it i.e.}
matrix configurations $X_i$ such that all differences between their
eigenvalues are large. In this situation all charged fields (relative
to the Higgs scalars) become very massive and effectively decouple from
the dynamics. Hence in this limit all matrix fields take the form
\eqn\diag{
\tilde{X}^I = \left(\matrix{~X^I_1~&&&\emptyset\cr &X^I_2 
&&&\cr %&& X^I_3 && \cr && \ddots &\cr
&&\ddots&\cr
\emptyset &&& X^I_N} \right) % \end{pmatrix}
}
and the corresponding effective SYM Hamiltonian reduces to 
the free field form
\eqn\ho{
H_0  = \hf \sum_{i=1}^N
\oint\! {d\sigma}\; \Bigl[ {1\over g_s^2}
E_i^2 + (\Pi^I_i)^2 +  (\partial_\sigma X^I_i)^2 + {\rm fermions}
\Bigr]}
where $E_i$ denotes the (diagonal part of the) electric flux.
% 
%\eqn\eflux{\tr E = e.}
%

For example, later in the paper we will consider matrices of the form
\eqn\genstate{{\vec X} = {\vecp\tau+ \vecb \over 2} 
\left(\matrix
{I_1/N_1&0\cr
0&-I_2/N_2\cr}\right)\ }
where $I_1$ and $I_2$ are $N_1 {\times} N_1$ and $N_2\times N_2$ identity
matrices.  This corresponds (see Appendix) to two widely separated
configurations with momenta
$p^+_{(i)} = N_{(i)}/R$ and $\vecp_{(i)} = \pm \vecp/2$, and with 
relative impact parameter $\vecb/2N_1 + \vecb/2N_2$.

In the SYM/IIA string dictionary, the electric flux $E_i$ gets
translated into D-particle number.  Indeed, in a canonical formalism the
electric flux is conjugate to the (zero mode of the) gauge field
$A_\sigma$, and thus represents the quantized momentum in the compactified
direction.  The quantum ground states of the SYM theory, corresponding to
asymptotic particle states, are thus labeled by their transverse momentum
$p_{\pperp}$, their light-cone momenta $p^+$ and $p^-$, and their D-particle
charge.

Depending on their topology, the eigenvalue fields $X^I_i(\sigma)$
in \diag\ combine into one or more separate strings. For example,
the trivial boundary condition 
\eqn\trivbc{X^I_i(\sigma+1) = X^I_i(\sigma)}
corresponds to a collection of elementary quanta (which may be thought of
as minimal length strings), whereas  
the string of maximal length $N$ is described by the periodicity 
condition
\eqn\boundaryc{
\qquad X_i^I(\sigma+1) = X_{i+1}^I(\sigma), \qquad i \in 
(1, \ldots, N).
}
We can write this condition as the $N \times N$ matrix equation
\eqn\perio{
X^I(\sigma+1) = V X^I(\sigma) V^{-1}
}
with $V$ the cyclic permutation matrix on the $N$ eigenvalues,
\eqn\cycle{V=
\left(\matrix{0~&~1~&&& \emptyset \ \ \cr &~0~&~1~&&\cr & %\emptyset 
&&\ddots &\cr &\emptyset &&&1 \ \cr ~1~&&&&0} \right).
}
As a result of this cyclic boundary condition, we can glue the
$N$ eigenvalue fields $X^I_i(\sigma)$ together into one 
single scalar field $X^I(\sigma)$ defined on a long interval 
$0\leq \sigma <  N$. Hence, when we
expand this field in the usual way in oscillation modes, the
frequency spacing between these modes is $N$ times smaller
than for a single valued field. 

In general the total matrix \diag\ will satisfy a 
periodicity condition of the form \perio\ with $V$ a block diagonal  
matrix consisting of (say) $s$ blocks of order $N_{(i)}$, such that
% \eqn{\sum_{i=1}^s N^{(i)} = N.} 
each block can be taken of the form \cycle, and thus (as described above)
defines a string of length $N_{(i)}$. As above, the space-time
interpretation of this length is as the light-cone momentum
\eqn\lcmom{
p_{(i)}^+ = \frac{N_{(i)}}{ R}\ .
}
This free string gas provides
a good description of the SYM Hilbert space in the limit where
all strings are far apart, {\it i.e.\ } if the eigenvalues in 
the matrix $X^I$ are well separated between the different blocks.

To avoid possible confusion later on, it is important to point out
that the matrix $V$ that specifies the periodicity condition on the
eigenvalues is {\it not} equal to the Wilson line of the gauge
field $A_\sigma$ around the $S^1$. This identification would arise
only if we insist on minimizing the potential energy term $(D_\sigma X)^2$ 
with respect to $A_\sigma$.

A related point is that for finite string coupling $g_s$, and certainly
in the large $g_s$ limit, the bound states with total light-cone momentum 
$N$ are clearly no longer necessarily described by means of a single
long string. More generally, one would expect that the bound state
wave function will have support on more subtle bound state configurations
consisting of several (up to $N$) short(er) strings. In other words,
it seems reasonable to expect that for large $g_s$ the long strings 
will tend to ``fractionate'' into many smaller constituents.

\subsec{D-particles and Electric Flux Sectors}

As we have indicated, an important new feature of the matrix string
formalism (relative to standard light-cone string theory) is that via the
electric flux, string states can also be adorned with a non-vanishing
D-particle charge.  In this subsection we will describe this
correspondence in somewhat more detail.  

To add to this interpretation, let us first show
that each separate string can carry only one type of electric flux.
Consider again the single string with length $N$.
Define the $U(N)$ matrix $U$ such that
\eqn\uv{
UV = VU e^{\frac{2\pi i}{ N}}
}%
with $V$ as in \cycle. Hence we can take
\eqn\formu{
U =\left(\matrix{1~&&& \emptyset \  \cr &e^{\frac{2\pi i}{N}}&&\cr
&&\ddots &\cr \ \emptyset &&&e^{\frac{2(N-1)\pi i}{N}}}\right)}
The $SU(N)$ part of the electric flux in this sector
is defined as
\eqn\hatu{
\hat{U} |\, \psi_e\, \rangle = \exp\Bigl(\frac{2\pi i e}{ N}\Bigr) 
|\,\psi_e\, \rangle}
with $e \in {\bf Z}_{N}$ and
$\hat{U}$ the quantum operator that implements the constant
gauge rotation 
\eqn\aix{
(A,X) \rightarrow (U A U^{-1}, UXU^{-1}).}

Since diagonal matrices are inert under this gauge rotation, 
we conclude that the $SU(N)$ part of the electric flux 
dynamically decouples from the diagonal configurations 
\diag\ that describe the separate freely propagating 
strings.
Now recall that in $U(N)$ SYM theory, the overall $U(1)$ part of the 
electric flux is related to the $SU(N)$ part $e$ via
\eqn\sunflux{\tr E = e \quad ( {\rm mod} \ N )\ .}
Supersymmetry ensures that the ground state in the $SU(N)$ sector
has zero energy even for $e \neq 0$. Hence the total ground state 
energy receives only a contribution from the overall $U(1)$ flux.
In the following we will thus identify $e$ with the total 
$U(1)$ electric flux. From the above description it is 
further clear that we can indeed turn on
only one electric flux per long string, as is appropriate for
its identification with D-particle charge.

The energy \ho\ of the ground state in this electric flux sector is equal to
\eqn\energy{
H_{0} =  \frac{e^2}{ 2 Ng_s^2}.}
General ground state configurations 
\eqn\grstate{
|N^{(i)},p^{(i)}_\pperp, e^{(i)} \rangle }
of $s$ separate strings of individual length $N^{(i)}$, 
transverse momenta $p_\pperp^{(i)}$, and D-particle charge 
$e^{(i)}$ have a SYM energy equal to
\eqn\hnul
{H_0 = \sum_{i=1}^s \frac{1}{ 2 N^{(i)}}
\Bigl[(p_\pperp^{(i)})^2 +  (e^{(i)}/g_s)^2\Bigr]} 
(recall this is defined with respect to the time $\tau=tR$)
which, when rescaled by $R$, is 
the sum of the $p_-$ light-cone momenta of the corresponding
collection of string ground states
\eqn\pminus{\sum_{i=1}^s \frac{1}{ 2 p_+^{(i)}}
\Bigl[(p_\pperp^{(i)})^2 +  (M^{(i)})^2\Bigr].}
In particular, we read off from 
\hnul\ that the states with D-particle charge $e^{(i)}$ each have 
mass  
\eqn\mass{
M^{(i)} = {e^{(i)}\over g_s}= {e^{(i)}\over R_9}}
in accordance with their identification as graviton states with non-zero 
KK momentum in the compact direction.

\subsec{IR limit and String Interactions}

In the IR limit, $g_s \rightarrow 0$, the SYM dynamics effectively
reduces to a free orbifold sigma model on the symmetric product
space $({\bf R}^8)^N/S_N$. The interacting theory for $g_s>0$ arises
by relaxing this IR limit. Correspondingly,
one may view the interacting string theory as obtained via a
perturbation of the $S_N$-orbifold conformal field theory.
In first order, this perturbation is described via a
modification of the CFT Hamiltonian
\eqn\twistop{
H = H_{0} + \lambda \int\! d\sigma \; V_{twist}}
Here $V_{twist}$ is an appropriate CFT twist operator that
generates simple transpositions of two string coordinates \refs{\DVV}.
In terms of the  bosonic twist-operators $\tau$ and fermionic
spin fields $\Sigma$ it takes the form
\eqn\Spintw{
V_{twist} =  \sum_{i< j} \left(\tau^I\Sigma_I \otimes
\overline{\tau}^J\overline{\Sigma}_J\right)_{ij}.}
This is a weight $({3\over 2},{3\over 2})$ conformal field.
The above twist field operator is an intertwiner between
different topological sectors of the orbifold model that are
related by a basic splitting and joining interaction between
two strings. Hence if we use the above Hamiltonian for computing
scattering amplitudes via standard perturbation theory, we will
indeed reproduce the conventional perturbation expansion of
type IIA string theory \refs{\ArFr}. This weak string coupling expansion
is a strong coupling expansion from the SYM perspective.

\newsec{Fixed Angle Scattering of Strings}

High energy, fixed angle processes in superstring theory were first 
studied in detail 
from the point of view of conventional string perturbation theory by 
Gross and Mende \refs{\GrMe}. Central to their approach is the observation 
that in the limit of large external momenta, the Polyakov path integral
at each given perturbative order is dominated by a finite number
of saddle point configurations. Furthermore, it was proposed 
that all these saddle points essentially describe the same preferred 
worldsheet trajectory, up to an overall factor depending on the 
loop order.

In the subsequent sections we will find independent evidence 
from the point of view of matrix string theory that supports this
physical picture. In this section we recall some
of the main results of \refs{\GrMe}. In addition we will 
give a useful characterization of the Gross-Mende saddle points
in terms of the light-cone gauge formulation of string perturbation
theory.

%%%%%%%%%%%%%%%%%%%%%%%%%%%%%%%%%%%%%%%%%%%%%%%%%%%%%%%%%%%%%%%%%

\ifig{\fig}{This figure indicates the kinematics of the transverse 
momenta $p_i$.}{\epsfysize=3.6cm\epsfbox{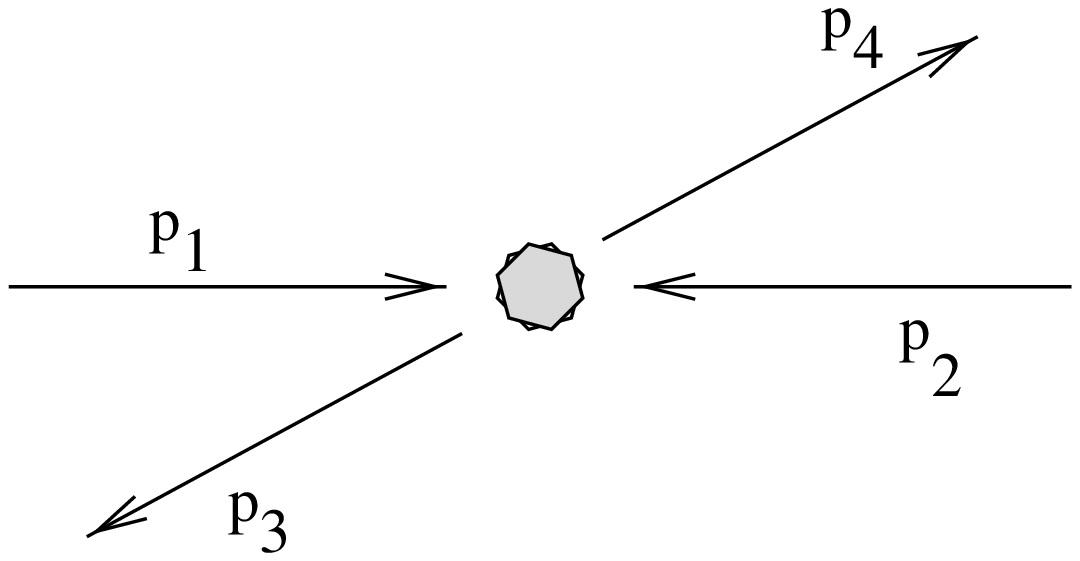}}

%%%%%%%%%%%%%%%%%%%%%%%%%%%%%%%%

It will be useful to first  recall a few geometric facts about the
light-cone gauge formulation of string perturbation theory.\foot{For
a review of some salient features, see \refs{\SBGTasi}.} 
Consider a tree level string diagram that
describes the scattering of four external massless particles with light-cone
momenta $p^+_i = N_i/R$ and transverse momenta $p_i$. 
For definiteness, we will describe this process in the center of mass 
frame in the transversal direction  
\eqn\cmframe{\eqalign{
{\vec p}_1 + {\vec p}_2 \is 0\cr
{\vec p}_3 + {\vec p}_4 \is 0}}
The four transversal momenta $p_i$ 
can all be chosen to lie within one given plane. We can thus write
all $p_i$ as complex numbers. In addition,
longitudinal momentum and energy conservation imply that 
\eqn\sumn{
N_1+N_2 = N_3 + N_4}
\eqn\sumpm{
{|p_1|^2\over N_1}+{|p_2|^2\over N_2}={|p_3|^2\over N_3}+{|p_4|^2\over N_4}}

For a given set of locations $z_i$ of the corresponding vertex operators,
the classical location of the worldsheet is described by  
\eqn\xplus{
X^+(z,\zbar) = \hf \sum_{i} \epsilon_i N_i \log|z-z_i|^2
}
\eqn\xtrans{
{X}(z,\zbar) = \hf \sum_{i} \epsilon_i {p}_i \log|z-z_i|^2
}
where $\epsilon_i = 1$ for the incoming and $-1$ for the outgoing
particles.
Here $(z,\zbar)$ denotes a general conformal parameterization of the
string world sheet. 

In the light-cone gauge, one chooses a fixed 
world-sheet parameterization by identifying $X^+$ with the world-sheet
time $\tau$
\eqn\lctime{
X^+(z,\zbar) =  \tau
}
which via \xplus\  amounts to setting
\eqn\lcgauge{
w\equiv \tau +i \sigma = {1\over 2\pi}  \sum_{i} \epsilon_i N_i \log(z-z_i)\ .
}
The differential $\omega=dw$ is a specific globally defined holomorphic 
differential on the world-surface; existence and uniqueness of such a 
differential at arbitrary genus\refs{\GiWo,\SBGTasi}
generalizes the construction to higher loop amplitudes.  
Notice that (due to the branch cuts in the logarithm) 
the coordinate $\sigma$ in \lcgauge\ is defined on an interval
$0\leq \sigma <  (N_1+N_2)$.

The light-cone coordinate system \lcgauge\ specifies a particular 
time-slicing
of the string world-sheet. In this coordinate frame, there are therefore
specific points on the world-sheet at which strings split or join. 
These interactions take place at zeros of $\omega$, that is 
critical points $z=z_0$ of the light-cone 
coordinate $X^+$, at which
\eqn\intpoints{
dX^+\mid_{z=z_0}=0\ .
%{\strut \Bigl{|}\textstyle{z=z_0}} = 0
}
Inserting the explicit form \xplus\ for $X^+$ gives
\eqn\zsolve{
\sum_{i=1}^s {\epsilon_i N_i \over z_0 - z_i} = 0 \ .
}
In the specific case of the four-point scattering amplitude, this condition
can be reduced to an equation relating the interaction point and the
cross ratio
\eqn\cross{
\lambda = {(z_1 - z_3) (z_2-z_4) \over (z_1 - z_2)(z_3-z_4)}
}
via
\eqn\znsolve{
{N_1\over z_0} + {N_2\over z_0-1} = {N_3\over z_0-\lambda}\ .
}
For given $\lambda$, this is a quadratic equation for $z_0$ with in 
general two solutions $z_0^+$ and $z_0^-$, representing the simple
splitting and joining interaction respectively.

Now we are ready to discuss the Gross-Mende saddle point. In the covariant
formulation used in \refs{\GrMe}, it is characterized by the condition
that it minimizes the Polyakov action for the classical trajectory
specified by \xplus-\xtrans. This action takes the form of a
two-dimensional ``Coulomb energy'' of four light-like ``charges'' given
by the momenta $p_i$:
\eqn\coulomb{
E_{C} = {1\over 2} \sum_{i<j} \; p_i\! \cdot \! p_j \; \log |z_i - z_j|^2\ .}
Due to conformal invariance, this energy $E_C$ depends on the 
locations $z_i$ of the vertex-operators only by means of the 
cross ratio $\lambda$. The variation of $E_C$ with respect to 
$\lambda$ reads
\eqn\ecvar{
\partial_\lambda E_C(\lambda) = {p_1\! \cdot\! p_3\over \lambda} + 
{p_2\! \cdot\! p_3\over \lambda-1}\ .}
The saddle point equation $\partial_\lambda E_C=0$ is solved by
\eqn\saddle{
\lambda = {p_1\! \cdot\! p_3\over {p_1\! \cdot\! p_2}} = {t\over s}\ .}
This saddle point corresponds to a particular classical world-sheet
trajectory which at high energies gives the dominant contribution to
the scattering amplitude.

%\eqn\sym{
%S = \frac{1}{ 2\pi } \int \!d\tau d\sigma\, \tr\left((D_\a X^I)^2 + 
%\theta^T \gamma^\a D_\a \theta +
%F_{\a\b}^2 - [X^I,X^J]^2 +\theta^T\gamma_I [X^I,\theta]\right).}

For later reference, it will be useful to translate the above description
of the GM saddle point into the light-cone gauge language.
To begin with, in the complex parameterization for the $p_i$, 
the Mandelstam parameters $s$ and $t$ are
expressed as 
\eqn\mandelst{\eqalign{
s \is - 2 p_1\! \cdot\! p_2 = (N_1 + N_2)^2{|p_1|^2\over N_1 N_2} \cr 
t \is - 2 p_1\! \cdot\! p_3 = {|N_3 p_1 - N_1 p_3|^2\over N_1 N_3} }}
so that \saddle\ takes the form
\eqn\saddl{
\lambda =  {N_2\over N_3} \, {|N_3 p_1 \!- N_1 p_3|^2\over  
(N_1+N_2)^2 |p_1|^2} .}
Together with \znsolve, this saddle point specifies a particular
set of locations for the two interaction points $z_0^\pm$ of the light-cone
string diagram. We now claim that this preferred location of the 
interaction points $z=z_0^\pm$ is singled out by the requirement that,
in the immediate neighborhood of $z=z_0^\pm$,
the transverse coordinate fields $X(z)$ are (anti-)analytic functions of $z$ 
\eqn\bpscond{\eqalign{
\partial_z X\mid_{z=z_0}
%{\strut \Bigl{|}{\textstyle z=z^+_0}} 
\is 0 \cr
\partial_\zbar \Xbar\mid_{z=z_0}
%{\strut \Bigl{|}{\textstyle z=z^+_0}} 
\is 0\ .}}

To verify this claim, let us compute the cross ratio $\lambda$ from \bpscond. 
The result should be equal to \saddle.  Inserting the solution \xtrans\ 
into \bpscond\ gives
\eqn\bpssum{
\sum_i {\epsilon_i p_i\over z^+_0-z_i} = 0}
In terms of the cross-ratio $\lambda$ defined in \cross\ this reads
\eqn\bpscross{
{p_1\over z^+_0(z^+_0-1)} =- {p_3 \over z^+_0 - \lambda}}
where we used that $p_1+p_2 = 0$.
When combined with the equation \znsolve, which relates $\lambda$ with 
the location of the interaction points $z_0$, this equation can  
indeed be used to compute $\lambda$ in terms of the scattering data. 
If we subtract $N_3$ times \bpscross\ from $p_3$ times \znsolve, we
obtain a linear equation for $z_0$, solved by
\eqn\znot{
z^+_0 = {N_1p_3 - N_3 p_1 \over (N_1+N_2) p_3}\ .}
%{N_1 \over N_1+N_2} + {p_1\over p_3} {N_3\over N_1+N_2}}
%
Further, from \bpscross\ we find that
\eqn\zznot{
\lambda  = z^+_0 \Bigl(1+ {p_3 \over p_1} (z^+_0-1)\Bigr)\ .}
After inserting \znot\ into \zznot, it's a simple calculation to
verify that the resulting expression for $\lambda$ indeed coincides with the
high energy saddle point \saddle.
Note that for the saddle-point configuration, $\lambda$ is in fact 
real. The interaction points $z^+_0$ and $z^-_0$ are in this case
each others complex conjugate.

In \refs{\GrMe}, Gross and Mende propose the following attractive 
generalization of the saddle-point to higher orders in the string 
perturbation expansion. They assume that the dominant saddle points 
at genus $G$ take the form of a $G+1$ fold cover of the same 
four-punctured sphere as described above, branched over the four locations 
$z_i$ of the vertex operators. The resulting surfaces are known
as $Z_N$ curves with $N=G+1$. The classical trajectory of these 
higher order saddle points has the same shape as the tree level 
trajectory, but 
its size is $N$ times smaller. (The intuitive reason is that they describe 
multiple wound strings,  so that the effective string tension is $N$ times
bigger than usual) . Correspondingly, since the different trajectories are 
weighted by the world sheet area, the higher order 
trajectories give contributions proportional to $e^{-E_C/N}$ (with $E_C$ 
given in \coulomb).  The higher genus contributions are thus quite strongly
enhanced at high energy.  We refer to \refs{\GrMe} for a more detailed 
description.

It is worth pointing out that the structure of the $Z_N$ curves and 
the corresponding space-time trajectories, as depicted in fig. 3,  
are quite reminiscent of the description of the ``long string'' boundary 
conditions in section 2. In our view, this (proposed) structure 
of the higher order interactions is one of several indications 
that the Gross-Mende approach to high energy string scattering may 
have a natural implementation in the Matrix string context.
The above light-cone characterization of the GM saddle point
in terms of the holomorphicity conditions \bpscond\ will be critical 
in establishing this relation!

\newsec{Matrix String Interactions}

In this section we will prepare the ingredients for the semi-classical 
study of high energy scattering in the Matrix string framework. 
To begin with, we notice that the above light-cone gauge description 
of the dominant string world sheet trajectories can rather easily be 
put into a matrix form, as follows. Starting from equations \xplus\ 
and \xtrans, we represent the classical string trajectory by means 
of a diagonal $N\times N$ matrix (with $N=N_1+N_2$) by first writing 
the transversal coordinates $\vec{X}$ as a function of $w$ defined 
in \lcgauge, and then ``roll up'' the spatial interval 
$0\!\leq \!\sigma\!<\! N$ onto the short interval 
$0\!\leq\! \sigma\! <\! 1$.
Concretely, we define the diagonal matrix elements of $X(\sigma)$
via $X_{kk}(\sigma) = X(\sigma+ k)$, and  in this way we indeed 
create matrix configurations that, away from the interaction times, 
satisfy the long string boundary condition \perio\ and \cycle.

These diagonal matrix configurations represent particular solutions
to the SYM equations of motion, that are regular everywhere {\it except}
at the interaction points.  If at some point in the $(\sigma,\tau)$ 
plane two eigenvalues $X_I$  and $X_J$ coincide, we enter a phase where 
locally the gauge symmetry is restored to $U(2)$. In general we should 
thus expect that in this local region the semi-classical
SYM solution will need to become truly non-abelian. 

It is readily seen that the diagonal matrix configurations constructed
via the above procedure from the CFT solution \xtrans\ is not single valued
around the interaction points. Instead, as explained in \refs{\DVV}, 
in going around the interaction point, the matrix $X$ undergoes a simple 
transposition of the two degenerating eigenvalues. 
In the gauge theory language, this means that the diagonal CFT
solution \xtrans\ in fact hides a delta-function Yang-Mills curvature
at the interaction point, such that the infinitesimal Wilson line 
around it coincides with this permutation group element \refs{\Wynt}. 
In this section we will describe how the Yang-Mills dynamics in fact
smoothes out this singularity.

%%%%%%%%%%%%%%%%%%%%%%%%%%%%%%%%%%%%%%%%%%%%%%%%%%%%%%%%%%%%%%%%%%%%%%%%%%

\ifig{\fig}{The string interaction relating a one string to a two string state.
This interaction occurs when two eigenvalues $X_I$ and $X_J$ coincide, we
enter a phase where an unbroken $U(2)$ symmetry is restored.}
{\epsfysize=3.3cm\epsfbox{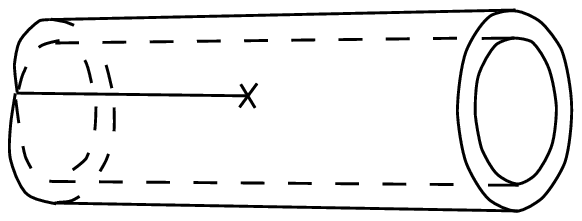}}

%%%%%%%%%%%%%%%%%%%%%%%%%%%%%%%%%%%%%%%%%%%%%%%%%%%%%%%%%%%%%%%%%%%%%%%%%%

Concretely, we will now exhibit a smooth and single-valued Yang-Mills 
configuration that describes the local splitting or joining of one or
two matrix strings. Ultimately, we will be interested in obtaining global
classical solutions to the SYM equations of motion that minimize the
Yang-Mills action for given asymptotic conditions on the matrix fields $X$,
as written in eqns \instate\ and \outstate\ in the Introduction.

\subsec{SYM Solution near Interaction Point}

It seems reasonable to assume that, at least in the immediate
neighborhood of the interaction point, these minimal action
configurations of the SYM model are described by supersymmetric
configurations. Hence, instead of trying to solve the full Yang-Mills
equations, we will restrict ourselves to the special class
of solutions satisfying a dimensionally reduced version of the 
self-duality equations from four to two dimensions.  
We will choose to work with
variables 
\eqn\cplxvar{X=\hf(X^1+iX^2)\ ,\ {\overline X} = \hf (X^1 -i X^2)\ ,}
setting the remaining $X^i$'s to zero.  The self-duality conditions then
become
\eqn\selfdual{\eqalign{
& F_{w\wbar} = - {i\over g^2_s}[\, X,\overline{X}\, ] \cr
& D_wX = 0 \cr
& D_\wbar\Xbar = 0 \ .
}}

The above equations are most conveniently analyzed by writing 
\eqn\Aglc{\eqalign{
A_w(w,\wbar) \is -iG \partial_w \Gi \cr
A_\wbar(w,\wbar) \is i   (\partial_\wbar \Gbar^{-1}) \Gbar }}
where $G(w,\wbar)$ denotes an element of the {\it complexified} (${\bar
G}\neq G^{-1}$) gauge group.
This parametrization of $A_\alpha$ allows one to solve the second and third
equation of \selfdual, via
\eqn\xsolve{
X(w,\wbar) = G \widehat{X}(\wbar) \Gi\ .  }
The first equation in \selfdual\ then takes the following form
\eqn\sinhG{
\partial_\wbar (\Omega \partial_w \Omega^{-1}) = -{1\over g_s^2}[ \Omega 
\widehat{X}(\wbar) \Omega^{-1}, 
\widehat{\Xbar}(w)]}
with 
\eqn\Omeg{\Omega = \Gbar G.}

Let us now look at the local neighborhood of an interaction point. 
For convenience, we choose coordinates such that it is located at $w=0$.
Since the interaction involves only two eigenvalues, it is sufficient
to consider only the corresponding $SU(2)$ part of the matrices.
The matrix $\widehat{X}$, which parametrizes the local
coordinate distance between the two interacting strings, can be 
chosen of the following form
\eqn\xparam
{\widehat{X}(\wbar) \simeq \pm B \; \sqrt{\, \wbar} \, \tau_3 }
for some constant $B$. The $\pm$ indicates that the interaction point $w=0$
represents a square root branch point for the diagonal matrix 
$\widehat{X}$ in \xparam, which therefore is multi-valued.

%In fact, if it where not for this multi-valuedness, the diagonal matrix
%$\widehat{X}(\wbar)$ would represent a perfectly valid solution 
%to the SYM equations near the interaction point. Instead, however,
%we must require that the physical matrix variable $X(w,\wbar)$ in \xsolve
%is single-valued around the interaction point, or alternatively,
%compensate its multi-valuedness by means of an appropriate boundary
%condition on the Yang-Mills potential $A_\alpha$ at $w=0$. We'll first 
%choose the latter option.
%

The diagonal matrix $\widehat X(\wbar)$, together with $A=0$, represents a
valid solution of the SYM equations \selfdual\ except at the interaction
point, where analyticity fails.  Therefore we'll look for a true solution
of the form \xsolve, where $G\rightarrow1$ asymptotically far from $w=0$.
A helpful {\it Ansatz} for $G(w,\wbar)$ is
\eqn\gparam{
G= e^{{1\over 2} \alpha \, \tau_1}  }
%
%\Gbar \is e^{-{1\over 2} \alpha \, \tau_1} }}
%
where for $\alpha(w,\wbar)$ we choose a real function (so 
that $G=\Gbar$ and $\Omega = \exp(\alpha \tau_1)$) 
that tends to zero far away from the interaction point.
We now compute
\eqn\comp{%\eqalign{
\Omega \widehat{X}\Omega^{-1} = B \sqrt{\wbar} e^{\alpha \tau_1}
\tau_3 e^{-\alpha\tau_1}
= B \sqrt{\wbar} 
\left(\matrix{\cosh 2\alpha & -\sinh 2\alpha\ \cr \sinh 2\alpha & 
-\cosh 2\alpha} \right).}
Hence
\eqn\comp{
\Bigl[ \Omega \widehat{X}\Omega^{-1}, \widehat{\Xbar}\Bigr] 
= 2|B|^2 |w| \, \sinh 2\alpha \; \tau_1}
and thus we find that under the present Ansatz the equation of motion 
\sinhG\ reduces to 
\eqn\sinhg{
\partial_w \partial_\wbar \alpha = {2\over g^2_s} |B|^2 |w| \sinh 2\alpha}
%+ i\pi \delta(w)}
%
which is essentially the familiar sinh-Gordon equation.  (It can
be transformed to the exact sinh-Gordon equation after a (multi-valued) 
coordinate transformation $w \rightarrow \tilde{w} = w^{3/2}$.)

The boundary condition that we must impose on $\alpha(w,\wbar)$ at $w=0$ 
follows from the requirement that the Yang-Mills configuration be regular. 
This condition is most easily understood in the gauge where $X$ is {\it
single-valued} near $w=0$; in this gauge the YM curvature
$F_{w\wbar}$ should be a regular function at $w=0$.  
The configuration
\xsolve-\gparam, however, is (for single-valued and real $\alpha$)
multi-valued. We can make $X$ single-valued by applying the
singular gauge transformation
\eqn\gaugetr{\eqalign{
X & \rightarrow U X U^{-1}\cr
A_w & \rightarrow - i U D_w U^{-1}}}
with gauge parameter
\eqn\utheta{
U = e^{\pm i\theta \tau_1/ 4}}
with $\theta = {1\over 2i} \log(w/ \wbar)$ the azimuthal angle around
$w=0$. In this gauge
\eqn\asing{\eqalign{
A_w \is i \Bigl[{1\over 2}\partial_w\alpha \pm {1\over 8w}\Bigr] 
\; \tau_1 \cr
A_\wbar \is \; -i \Bigl[{1\over 2}\partial_\wbar\alpha \pm {1\over 8 \wbar}
\Bigr] \; \tau_1}}
Using that $\partial_w {1\over \wbar} = \pi \delta^{(2)}(w)$, this gives
\eqn\fww{
F_{w\wbar} = -i\tau_1 \Bigl(\partial_w \partial_\wbar \alpha \pm {\pi \over 4} 
\delta^{(2)}(w)\Bigr).}
The regularity requirement at $w=0$ is therefore that
$\partial_w \partial_\wbar \alpha \simeq \mp {\pi \over 4}
\delta^{(2)}(w)$. We thus deduce that the solution to equation 
\sinhg\ that we want must satisfy the following asymptotic condition
\eqn\alphlimit{\qquad \qquad 
{\alpha}(w,\wbar)  \; \simeq \;  \mp{1\over 2} \log | w| + {const.}  
\qquad \quad 
w \rightarrow 0}
while at large distances from the interaction point ${\alpha}$ must tend to 
zero.

Now let us write $\alpha = \alpha(r)$ with $r = |w|$. The equation
of motion \sinhg\ reduces to the ordinary non-linear differential
equation 
\eqn\sing{
(\partial_r^2 + {1\over r} \partial_r) {\alpha} = 
{8\over g^2_s} |B|^2 r \sinh 2\alpha\ .}
%
%Note that the interaction point gets mapped to $r \rightarrow - \infty$.
A numerical solution to this equation is depicted in fig. 6.
For large $r=|w|$ the solution looks like 
\eqn\rlimitp{%\eqalign{
\alpha(r)  \sim  %J_0({|b|\over 3 g_s} (2r)^{3/2}) \cr
{\mp 1\over |w|^{3/4}} \exp\Bigl(-{8|B|\over 3 g_s} 
|w|^{3/2} \Bigr)\ .}
%

%%%%%%%%%%%%%%%%%%%%%%%%%%%%%%%%%%%%%%%%%%%%%%%%%%%%%%%%%%%%%%%%%

\ifig{\fig}{The numerical solution $\alpha(s)$ to $(\partial_s^2 + 
{1\over s} \partial_s) {\alpha} = s \sinh 2\alpha$, as 
a function of the distance $s$ from the interaction 
point. The initial condition for small $s$, that after integrating 
leads to the correct asymptotic behavior for large $s$, reads 
$\alpha(s) = -{1\over 2} \log s + 0.0305070(1)$.}
{\epsfxsize=5cm\epsfbox{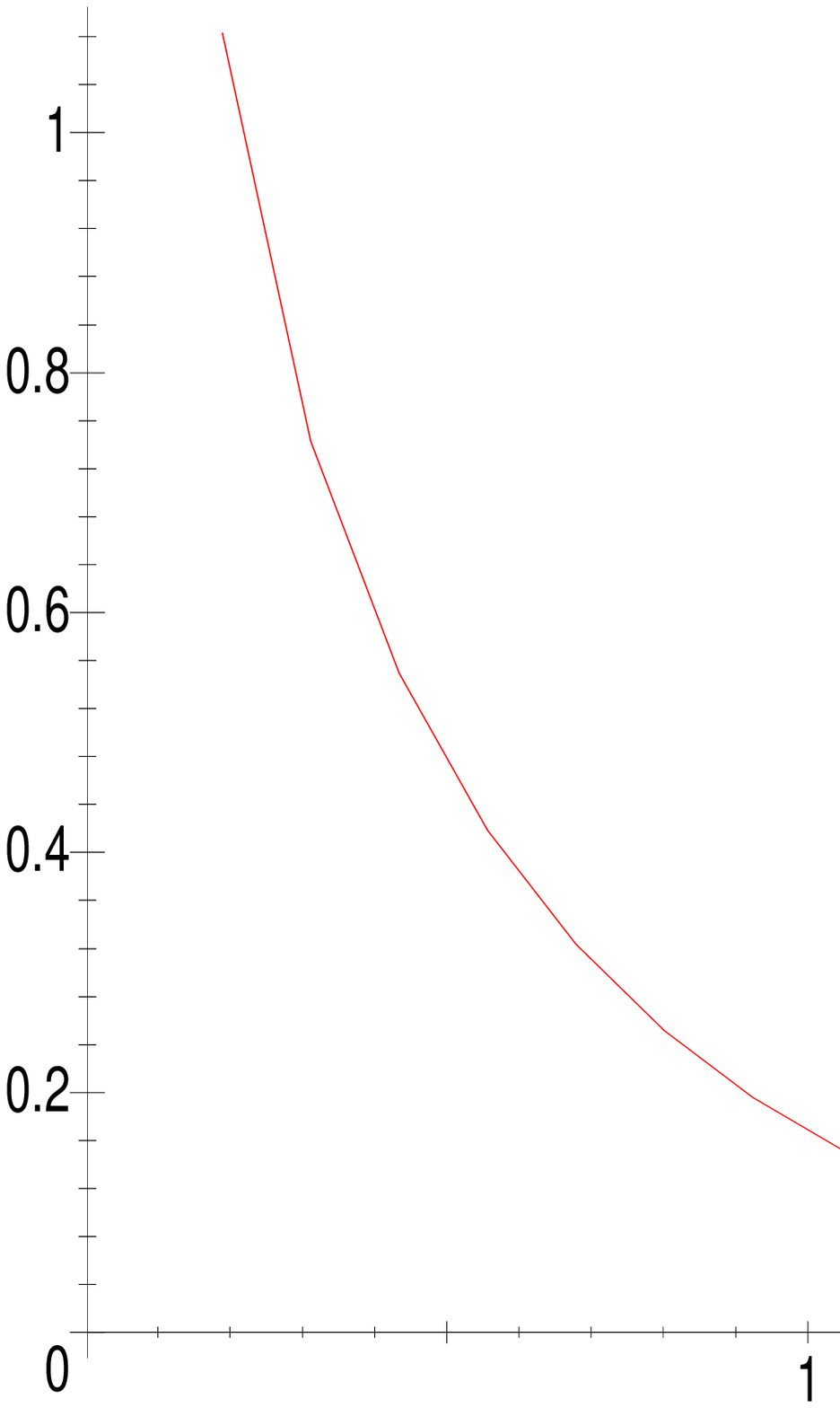}}

%%%%%%%%%%%%%%%%%%%%%%%%%%%%%%%%

\newsec{High Energy Scattering of Matrix Strings}

The above matrix solution of the string interaction should be
viewed as a local description in the immediate neighborhood
of the interaction point. In general, it must therefore be glued
via an appropriate patching procedure into a complementary
CFT type solution ({\it e.g.} as described in section 3) that matches 
with the asymptotic scattering data at the far past and future. 
The idea here is that (as we will see shortly) at sufficiently 
high collision energies, the size of the interaction regions are 
small compared to the rest of the matrix string world sheet. 
Hence, while the behavior \xparam\ provides the asymptotics 
for large $|w|$ at the UV scale of the matrix solution, it also provides 
the local boundary condition near the interaction point for the
CFT solution for $X$ that describes the IR part of the saddle point.

The solution \selfdual\ that we described is  not the most
general SYM description of a string splitting and joining event,
but rather the most symmetric one, with smallest action.
This means therefore that there is a non-trivial matching
condition on the corresponding matrix string world sheet: 
from \xparam\ we see that we must require that the transverse
string coordinates $X$ behave holomorphically near the 
interaction point. Remarkably, this is exactly the same  
condition  as \bpscond, which tells us that the shape of the
string world sheet must be precisely that of the 
Gross-Mende saddle point!  Therefore these solutions indeed 
seem appropriate to a YM generalization of 
the high-energy scattering of \GrMe. In this section, we fill in 
a few more details of this connection.

\subsec{Evaluation of the classical action}

In order to estimate scattering amplitudes via the instanton processes, one
must calculate the instanton action.  The bosonic part of the SYM action 
(with only two $X$-fields non-vanishing) 
can be written as
\eqn\bpsact{\eqalign{
S = \int& d^2 w \left\{-{g_s^2} 
\Bigl(F_{w\wbar} + {i\over g^2_s}[\, X,\overline{X}\, ]\Bigr)^2
+ 4 D_w X   D_\wbar\Xbar\right\}\cr 
&+\oint (\Xbar {\overline D} X + X D \Xbar - \Xbar DX - X {\overline D}\Xbar )
}} 
and thus for the supersymmetric configuration that satisfy
\selfdual, the total classical action reduces to a boundary term
\eqn\bounds{
S_{cl} = 
\oint (\Xbar {\overline \partial} X + X \partial \Xbar)}
identical to the boundary term needed to glue the non-abelian
matrix solution described in the previous section into the
CFT type solution.
Hence we claim that, in the limit that the matrix interaction points
become sufficiently small, the SYM action for the above saddle point
configurations coincides with the CFT action, {\it i.e.} for the case of
a tree level string diagram it equals the ``Coulomb energy'' \coulomb,
where me must insert the saddle-point value for locations $z_i$.
It is perhaps worth pointing out that this saddle point actions is 
indeed fully Lorentz invariant, as it should be. While this is not 
surprising once we've established the connection with the GM saddle 
point, it does seem to represent a rather non-trivial statement from 
the SYM point of view!

More generally we see that from \bpsact\ we can derive
(as usual) an inequality, which suggests that whenever the interaction
does not take place at a holomorphic point for the $X$-fields,
the SYM action is always larger than the corresponding CFT
action. This provides additional evidence for the conjecture that 
the above type of configurations indeed represent dominant saddle-points,
that minimize the SYM action. 

Obviously, there exist a large number of CFT-type solutions for which
$X$ varies (anti-) holomorphically near all interaction points. In
particular, there are the higher genus $Z_N$-curves of \refs{\GrMe}.
In addition it is also possible to write down SYM solutions that 
describe multiple string world sheets, but nonetheless still satisfy
the appropriate boundary conditions, as specified in eqns
\instate\ and \outstate\ in the Introduction. Ideally, one would
like to know which (sub-class) of these solutions provide the truly
dominant contribution to the scattering amplitude. We will not attempt
to answer this question here, and will restrict ourselves to some 
qualitative remarks in the concluding section.

\subsec{Minimal distance}

The parameter $|B|$ that governs the size
of the interaction vertex, as seen in \rlimitp, can be straightforwardly
determined in terms of the momenta of the external states.
The coordinate system $(w,\wbar)$ on the Yang-Mills cylinder that we used
in the analysis of the self-dual Yang-Mills equation \selfdual, coincides 
with the light-cone coordinates defined in \lcgauge. From this and \xparam\
we immediately find 
\eqn\bexpr{
|B|^2 \, = \,2 
{\;\; |\partial_\zbar X(z_0^+)|^2\over |\partial^2_z X^+(z_0^+)|}\ .}
A straightforward calculation then gives 
\eqn\bans{
|B|^2= {|p_1\overline{p}_3-
\overline{p}_1 p_3| (N_1+N_2)\over \sqrt{N_1N_2N_3N_4}}\ .}

It is interesting to note that for this
solution, even though the eigenvalues of the complex coordinate matrix
$X$ indeed vanish at the 
interaction point, the full matrix coordinate $X$ in fact does not! 
Instead, near $w=0$ it approaches the constant non-diagonalizable
matrix 
\eqn\xlimit{
\qquad \qquad 
X(w,\wbar)  \; \simeq \;  const. \ g_s^{1/3} B^{2/3}  
\left(\matrix{1 & \mp 1 \cr \pm 1 & -1 } \right)
 \qquad \quad 
w \rightarrow 0}
(The value of the overall constant can quite easily
be determined numerically.) 
{}From this we read off that the minimal ``distance'' between the 
two interacting strings is in fact non-zero! Instead, we have
\eqn\dlimit{
d_{min} = \sqrt{\tr(X(0) \Xbar(0))} \sim  g_s^{1/3} |B|^{2/3}.}
Although it is tempting to speculate (as indeed we will do in the concluding
section), the precise physical significance of this result is as yet 
unclear to us.  We do notice, however, that the typical world sheet 
size $\ell_{inst}$ of the matrix interaction region, 
as can be read off from \rlimitp, 
is naturally expressed in terms of this minimal relative distance as 
$\ell_{inst} = (g_s/|B|)^{2/3} = g_s / d_{min}$.

\subsec{Fluctuation Determinant}

In principle it should be possible to compute the one-loop determinant
of the quantum fluctuations around these semi-classical configurations.
An important motivation for performing such an analysis is 
to obtain a semi-classical estimate for the absolute strength
of the splitting and joining interactions in matrix string theory.
Duality symmetries of M-theory gives the precise prediction that 
this strength should be governed by the string coupling $g_s$. 
To verify this, one would need to compare the SYM one-loop determinant
with the Gross-Mende fluctuation determinant, coming from the
gaussian integration over the Riemann surface moduli around the
saddle-point. We leave this for future work. 

It seems even more worthwhile to look for true new physical
effects that might arise from the one-loop corrections. Compared to
the conventional perturbative string description, the new degrees
of freedom in matrix string theory are the charged components of 
the $X$-fields, as well as the extra gauge potential $A_\alpha$. 
These new degrees of freedom are non-perturbative from the string 
perspective, and their quantum fluctuations could thus potentially 
lead to new physics. As we will show in the next section, there
is indeed such a new effect: the pair creation of D-particles.

\newsec{D-Particle Pair Production} 

In this section we turn to the process of pair creation of D charge, which
is in our description $x^9$ momentum or equivalently (under the Matrix
string duality) electric flux.  This can be viewed as a contribution to the
fluctuations about the high-energy scattering processes of the preceding
sections, or as a process worthy of interest in its own right in the
context of graviton scattering.
There are several viewpoints from which
this can be investigated.  In the limit where $x^9$ decompactifies, this
simply matches onto the standard supergravity calculation\refs{\BeCo}.  
In fact, we can
work backwards from this, using the method of images, and compute the
amplitude at large finite $R_9$, in the special situation with source and probe
particles, $N_1\gg N_2$.  We will discuss this calculation first.  
Alternately, one can study this process directly
in the matrix string approach, and derive the pair-production rate via a
one-loop Yang Mills calculation.  This latter approach gives a leading
order result valid for arbitrary $N_1$ and $N_2$, and also more readily
makes connection with the other results of this paper.  Furthermore, the
Yang-Mills calculation also apparently extends beyond the region
where supergravity is a valid approximation.

\subsec{Supergravity calculation}

Consider 11-dimensional supergravity compactified on $S^1\times S^1$,
where one $S^1$ a lightlike circle of
``radius'' $R$
\eqn\dlcq{x^- \equiv x^- + 4\pi R}
and the other $S^1$ denotes a space-like circle of radius $R_9$. 
As we've seen,  in the M-theory/matrix string correspondence this second radius
$R_9$ is expressed as  $R_9 = g_s $ in string units.

Consider in this set-up the scattering process of two massless
particles of light-cone momenta $p^+_i = N_i/R$ and transverse momenta $p_i$. 
Let us first consider the probe situation $N_2\ll N_1$. Then one
can already get quite useful information about the scattering process
from considering the classical gravitational force between the two
particles.
The boosted particle with $p_+=N_1/R$ produces via its stress-energy 
a non-trivial gravitational background, described by the generalized 
Aichelburg-Sexl shockwave geometry of the form \refs{\AiSe,\BBPT}
\eqn\ASm{ds^{2}=-dx^{-}( dx^{+}+f(r,a)dx^{-}) +dx_{\bot }^{2} +g_s^2 da^2.}
with
\eqn\fr{f(r,a) = 
\sum_{k} {15 N_1 g_s^3 \over 2 R^2 
(r^2 +g_s^2 (2\pi k + a)^2)^{7/2}}}
Here $a$ denotes the coordinate distance from the gravitational source in the
compact $x^9$ direction, and the sum over $k$ arises from the image points 
in this direction.
 
The momentum four vector
of the second massless particle moving in this background geometry 
will satisfy a dispersion relation of the form
\eqn\dispersion{2 p^-(p^+ + f(r,a) p^-) = p^2 + e^2/g_s^2 }
where $e$ denotes the quantized momentum in the $x_9$ direction.
We can solve for the light-cone hamiltonian $p^-$ of the
particle and obtain
\eqn\pminus{p^- = 
{p^+\over 2 f(r,a)}\left\{\sqrt{\Bigl(1 - {2 f(r,a)\over (p^+)^2} 
(p^2+ e^2 /g_s^2)\Bigr)} -1\right\} }
Substituting $p^+ = N_2/R$, (and rescaling the light-cone time by a factor 
of $R$) we can write this as 
\eqn\htwo{H = H_0 + H_{int}}
where 
\eqn\hnot{H_0 = {1\over 2N_2} (p^2+ e^2/ g_s^2)}
and
\eqn\Hint{H_{int}
\simeq  - {15 N_1 g_s^3 \over 8 N_2^3} \sum_{k} {(p^2+e^2/g_s^2)^2 \over 
\Bigl(r^2 +g_s^2 ( 2\pi k  +  a)^2\Bigr)^{7/2}} + \ldots}
Hence the motion of the second particle in terms of the light-cone
time $x^+$ looks like that of a particle with mass $N_2$ moving
in $R^8 \times S^1$ under the influence of an interaction potential
given by \Hint. 

{}From this description we can now quite easily extract a low
energy prediction for the D-pair production rate. To this end,
it is useful to rewrite the interaction Hamiltonian via a
Poisson resummation as
\eqn\hint{
H_{int}\simeq
-{1 \over 2\pi} {N_1 \over N_2^3}
g_s^2(p^2+e^2/g_s^2)^2
\sum_n \exp(ina) \int dT T^2 \exp(-Tr^2) \exp(- n^2/4g_s^2T)
}
The $n=1$ term in this series is the term that corresponds to changing the
compact momentum by one unit, \ie\ to D-charge production.  Working to
first order in perturbation theory, we can then compute the corresponding
phase shift, using 
\eqn\phham{\delta=-\int d\tau H_{int}(b^2 +p^2\tau^2)\ .}

\subsec{D-pair production via electric flux creation} 

We now study this problem of D-charge creation in the Matrix string
framework.  
More generally, we consider scattering 
states which 
asymptotically have momenta of the
form 
\eqn\momdef{p^\mu=(p^-,\vecp,p_9=n/R_9,p^+=N/R)\ .}
These include both gravitons ($n=0$) and D0-branes -- or anti-branes --
($n=\pm1$).  The case of current interest begins with an initial state of
two gravitons, and pair produces a D particle pair.  This process is
intrinsically non-perturbative from the point of view of string theory.  It
is also a process not accessible in the standard Matrix theory approach,
where the anti-branes are boosted away to infinite energy.  

In principle (for example on a sufficiently large computer) 
it appears possible to calculate such amplitudes to arbitrary
order in the coupling $g=g_{YM}=1/g_s$, and calculate the D-pair production
rate even for small $g_s$.
In this section we will work to leading
non-trivial order (one-loop), and leave further calculations to other
work.  Similar calculations have been performed in the context of Matrix
theory in \refs{\BeBe}.

\subsec{One-loop calculation, $N=2$}

For simplicity we begin with the case where the incoming and
outgoing particles all have $N=1$.  The next subsection will
generalize to arbitrary $N$.  The asymptotic states thus take the form
\genstate; specifically,
\eqn\initstate{{\bar X}^1=\hf\left(\matrix{p\tau&0\cr0&-p\tau}\right)\ ,\
{\bar X}^2 =
\hf \left(\matrix{b    &0\cr0&-b}\right)\ ,}
corresponding to two particles with center of mass momentum $p$ and impact
parameter $b$ (measured in string units).
It will also be 
useful work with a non-trivial gauge background 
\eqn\clfield{
{\bar A}_\sigma=\left(\matrix {a/2+e\tau/g_s^2&0\cr 0&-a/2 -
e\tau/g_s^2\cr}\right) ,\ {\bar E }= \left(\matrix
{e&0\cr 0&-e\cr}\right)\ .}
The constant electric field corresponds to a non-zero D-charge
for the incoming and outgoing particles, with quantization
\eqn\equant{e \in {\bbb Z}\ .}
The prototypical example of production of D-charge is 
in processes where this
changes by one unit,
\eqn\elflux{\Delta E=\left(\matrix{1&0\cr0&-1}\right)\ .}
Introducing the constant background potential $a$ will help keep track of
such changes.

In the large string coupling/small Yang-Mills coupling limit, the
leading contribution to D-charge producing processes is easily computed 
via a one-loop super-Yang Mills calculation.
Calculations at higher loop order then give subleading corrections
in $g = 1/{g_s}$.

Our starting point is the Yang-Mills action \finact, although it will
be simpler to begin with it written in its un-dimensionally reduced
form in terms of the gauge field
\eqn\Adecomp{
A^M=(A^\mu,gX^i),}
with ${M=0}, \cdots ,{9}$, ${\mu} = {0}, {9}$ and ${i} = {1},
{\cdots}, {8}$.  We will decompose the gauge field into background and
fluctuation pieces,
\eqn\backdecomp{
{A_M} = {\bar A}_M + {g} {\tilde A}_M\ .}

The Feynman background gauge-fixed lagrangian is
\eqn\gflang{
{\cal L} = {-}{\rm Tr} \left\{ {{1}\over{4g^2}} {Tr} ({F_{MN}^2}) {+} {{1}\over2{g^2}}
{Tr}({\bar D}^M{A_M})^2 {-} {i}{\bar\psi}{\qi}{\psi}\right\}    }
\noindent where ${\bar D}_M = {\partial_M} + {i}{\bar A}_M$.
Using the decomposition \backdecomp\
\noindent we find
\eqn\sdecomp{\eqalign{{\cal L}&=-Tr \Biggl\lbrace{{{\bar F}^2}\over{4g^2}} +\hf ({\bar
D}_M {\tilde A}_N)^2 + {i}{\bar F}^{MN} [{\tilde A}_M,{\tilde A}_N] {-}
{i}{\bar\psi}{\qi}{\psi}\cr &+{g} {\bar\psi} {\q}{\psi}+{ig}{\bar
D}_M{\tilde A}_N [{\tilde A}^M,{\tilde A}^N] {-} {{g^2}\over{4}} [{\tilde
A}_M, {\tilde A}_N]^2\Biggr\rbrace.\cr}}

The amplitude in question is given by
\eqn\fetint{
{\calA} (a,e) =  \int {\cal D}
{\tilde A}_\mu{\cal D}{\tilde X}^i{\cal D}{\psi}{e^{iS}}}
where the boundary conditions on the functional integral are
chosen to correspond to the asymptotic behavior given in
\initstate,\clfield.   

If we write
\eqn\Asum{
{A_M} = {{1}\over{2}} {A_{Ma}}{\sigma}^a = {{1}\over{2}} {A_M}_+
{\sigma}^+ + {{1}\over{2}} {A_M}_- {\sigma}^- {+}\hf {A_{M3}}
{\sigma}^3,}
\noindent with
\eqn\sigmadef{
{\sigma}^{\pm} = {{{\sigma}^1 \pm {i}{\sigma}^2}\over{\sqrt{2}}},}
\noindent then the couplings in \sdecomp\ include the
standard charged minimal couplings of ${A_+}$, ${A_-}$, ${\psi_+}$, and
${\psi_-}$ to the U(1) field ${\tilde A}_{\mu 3}$.  The amplitude to create
unit electric flux is therefore given by summing the loop contributions to
\fetint\ in which one of these charged particles circulates once around the
${\sigma}$-direction; higher encirclings yield more flux.  Therefore we
need the contribution of the charged state windings to the one-loop amplitude.

This immediately follows by reading off the spectrum
from the second through fourth terms of \sdecomp\ in the backgrounds 
\initstate\ and \clfield.  We begin by defining 
\eqn\phatdef{\phat^2 = g^2 p^2 + 4 g^4 e^2\ .}
In the bosonic sector we find the neutral
fields
\eqn\neutf{\eqalign{
{\tilde X}_i^3,&\ i=1,\cdots,8;\quad m^2=0\cr
A_\mu^3,&~~~~~~~~~~~~~~~~~~~~~m^2=0\cr}}
\noindent and the charged fields
\eqn\bosmass{
\eqalign{
{\tilde X}^\pm_i\, ,\,i=2,\cdots,8\quad;&\quad m^2=r^2\equiv \phat^2 
\tau^2 + a^2 + g^2b^2\cr
{{1}\over
%{\sqrt{\ehat^2+p^2}}
\phat} (gp {\tilde A}^{9\pm} {-} 2eg^2 {\tilde
X}^{1\pm})\quad;&\quad m^2=r^2\cr
{\tilde A}^{0\pm} + {{i}\over %{\sqrt{\ehat^2+p^2}}
\phat} (2eg^2 {\tilde A}^{9\pm} + gp
{\tilde X}^{1\pm})\quad;&\quad m^2=r^2+2 %{\sqrt{p^2+\ehat^2}}
\phat\cr
{\tilde A}^{0\pm} - {{i}\over %{\sqrt{\ehat^2+p^2}}
\phat} (2eg^2 {\tilde A}^{9\pm} + gp
{\tilde X}^{1\pm})\quad;&\quad m^2=r^2-2 \phat %{\sqrt{p^2+\ehat^2}}
.\cr}}

For the fermions, we have 16 massles uncharged states, 8 charged states
with masses $m^2 = r^2 + %\sqrt{p^2+\ehat^2}
\phat$, and 8 charged states with masses $m^2 =
r^2- %\sqrt{p^2+\ehat^2}
\phat$, as in \refs{\BeBe}. Finally, including the ghosts gives one
complex, uncharged field ${C^3}$ with $m^2=0$ and one complex charged
field ${C^\pm}$ with $m^2=r^2$.

All of the charged fields are minimally coupled to the background field
${\bar A}_9 \equiv {\bar A}_\sigma$. 
At one loop level, we have
\eqn\oneloopamp{
\calA_1 \rm {(a,e)} =  \int \prod_I {\cal D}{\Phi_I}e^{iS^{(2)}[{\Phi_I}]}}
where I labels the charged fields enumerated above (the
uncharged contributions cancel), arbitrary winding is allowed, and where
$S^{(2)}$ is the quadratic part of the action \sdecomp\ including the
coupling to ${\bar A}^3_\sigma$ through ${\bar D}$. Working with phase
shifts, we then have
\eqn\lnamp
{\eqalign{i\delta_1&= {ln} {\calA_1} = \sum_I {ln} \sum_n \int_n
{\cal D}{\Phi_I} e^{iS^{(2)}[\Phi_I]}\cr
&=- {\sum_I} (-{1})^{F_I}{\sum_n}{\int_0^\infty}{{dS}\over{S}}
\int_n{\cal D}{\tau}{\cal D}{\sigma}\,{e^{i\int_0^S ds
\bigl\lbrack{\dot\sigma}^{\mu2}/2 - {\bar A}^3_\mu {\dot\sigma}^\mu
({s}) - {{m^2_I(\tau)}\over{2}}\bigr\rbrack}}.\cr}}
Here we have used the functional integral representation in
terms of the first-quantized trajectory ${\sigma^\mu}({s}) = ({\tau}
({s}),{\sigma}({s}))$, $n$ is the winding number about the cylinder, 
and ${F_I}$ denotes fermion number of the
field. 

For general winding ${n}$ the functional integrals in \lnamp\ are readily
rewritten in terms of functional determinants. For example, with
${m^2}={r^2}$ we have
\eqn\Pintrg{
\eqalign{
&{\int_n}{\cal D}{\tau}{\cal D}{\sigma}\,{e^{i\int_0^S ds\bigl\lbrack
{\dot\sigma}^{\mu2}/2 - g^2(p^2\tau^2+b^2)/2 -
{\bar A}_\sigma^3{\dot\sigma}({s})\bigr\rbrack}}\cr
&=e^{- ina+i n^2/2S-ig^2b^2S/2}\Delta(p,e,S)}}
where
\eqn\Detexp{
\Delta(p,e,S) = {\rm
det}^{-1/2}\left(\matrix{\partial^2_s-g^2p^2&-2g^2e\partial_s\cr
2g^2e\partial_s&-\partial_s^2}\right)\ .}
\noindent Combining such expressions and defining $S=2T$ then gives
\eqn\pephase{
i\delta_1(a,p,e)=\sum_n\,{e^{- ina}}\int_0^\infty {{dT}\over{T}}
{e^{in^2/4T-ig^2b^2T}}\Delta(p,e,T)
\biggl\lbrack-6-2{\cos}(2gp T)
+8{\cos}
({gp T})\biggr\rbrack.}
Recalling the quantization rule \equant, we see that as long
as $gp\gg 1$ the electric
background only contributes at higher order in $g$; neglecting this, the
determinant is readily evaluated using 
\eqn\deteval{
\eqalign{
{\rm det}^{{1}\over{2}}i(-\partial^2_s+g^2p^2)&={gp}\,{\prod^\infty_{n=1}}
\Biggl\lbrack\Biggl({{2n\pi}\over{S}}\Biggr)^2 +g^2p^2\Biggr\rbrack =
{2}
{\sinh}\left({gpT}\right)\cr
{\rm det}^{{1}\over{2}}i(-\partial^2_s)&={\sqrt{2\pi T\over i}}\ .\cr}}
The  phase shift then becomes
\eqn\onephase{\eqalign{
i\delta_1(a,p,e)=\hf\sqrt{i\over 2\pi}\sum_n\,e^{- 
ina}\int_0^\infty {{dT}\over{T}}&
{e^{in^2/4T-ig^2b^2T}}
{1\over \sqrt{T} \sinh(gpT)}\cr
&\biggl\lbrack-6-2{\cos}(2gpT)
+8{\cos}
({gpT})\biggr\rbrack. 
}}

In the full amplitude $\calA_1=\exp\{i\delta_1\}$, the coefficient of the term
$e^{- ina}$ is the amplitude to make a transition from a state with electric
flux $e$ to $e+n$:
\eqn\ampdecomp{\calA_1(a,e)= \sum_n e^{- ina} \langle e+n|e\rangle\ .}
We have found that this is independent of $e$ to order $g^2$. The
amplitude for a change by one unit of charge (\eg\ two gravitons to
$D{\bar D}$ pair), as well as the effective interaction hamiltonian, can be
derived from these expressions in the range
$p\ll b^2$.  There the integrand in \onephase\ can be expanded in $pT$ 
to find
\eqn\phasap{i\delta_1(a,p,e)\approx - {g^3p^3\over 2} \sqrt{i\over \pi} 
\sum_n e^{- ina} \int {dT} T^{3/2} e^{in^2/4T -ig^2b^2T}\ ;}
the leading order $D{\bar D}$ production amplitude is just the coefficient
of $e^{- ia}$ in the series.  From
\phasap\ and \phham\ we can also work backwards to extract the effective
hamiltonian.  We find
\eqn\phasham{ H_{int}= -{15\over 8} g^{-3}p^4\sum_k {1\over 
\left[r^2 + (a+2\pi k)^2/g^2\right]^{7/2}}\ ,}
in agreement with \Hint.

\subsec{Generalization to arbitrary ${N}$}

The one-loop calculation of the preceding subsection is readily
generalized to the case where the incoming and outgoing particles have
arbitrary ${p_{11}}$, or equivalently, ${N}$. In this case there are a
variety of different boundary conditions that may be placed on the
${N}{\times}{N}$ blocks. Two that we will consider are the trivial
boundary condition,
\eqn\trivbc{
X (\sigma + 1) = X (\sigma)\ ,}
and the single long string boundary condition,
\eqn\stringbc{X (\sigma + 1) = V^{-1} X(\sigma) V}
where ${V}$ is given in \cycle.

In the case of two incoming states with momenta $N_1,N_2,$ we write
\eqn\Xexpan{
X^i={\bar X}^i + {\tilde X}^i}
\noindent where $X^i$ is an $(N_1+N_2){\times}(N_1+N_2)$ matrix. In
particular, the background is taken to be
\eqn\Xbackg{
\eqalign{
{\bar X}^1&={{p\tau}\over{2}}\left\lgroup\matrix
{I/N_1&0\cr
0&-I/N_2\cr}\right\rgroup
\equiv {{p\tau}\over{2}}\,T_D\cr
{\bar{X^2}}&={{b}\over{2}}\,T_D\cr}}
where we have split the matrix into $N_1 {\times} N_1$ and
$N_2\times N_2$ blocks corresponding to two ``clusters,'' and $I$
represents the corresponding identity matrices.

A useful decomposition of the fluctuations ${\tilde X}^i$ is in terms of
the matrices
\eqn\genmatr{
T^{a_1}=\left\lgroup\matrix{t^{a_1}&0\cr
0&0\cr}\right\rgroup,\ 
T^{a_2}=\left\lgroup\matrix{0&0\cr
0&t^{a_2}\cr}\right\rgroup,}
where ${t^{a_i}}$ are hermitian generators of $SU(N_i);
T^{\alpha_1\alpha_2}_+, T^{\alpha_1\alpha_2}_-$, which have matrix
elements
\eqn\Matrdefs{
\eqalign{
(T^{\alpha_1\alpha_2}_1)_{\beta_1\beta_2}&={\sqrt{2}}\delta_{\alpha_1\beta_1}
\delta_{N_1+\alpha_2\beta_2}\cr
(T_2^{\alpha_1\alpha_2})_{\beta_1\beta_2}&={\sqrt{2}}
\delta_{N_1+\alpha_1\beta_1}
\delta_{\alpha_2\beta_2};\cr} }
and $T_D$:
\eqn\XTexpan{
{\tilde X}^i= {{\tilde X}_D\over{2}}T_D + {{\tilde X}_{a_1}\over{2}}T^{a_1} +
{{\tilde X}_{a_2}\over{2}}T^{a_2} + 
{{\tilde X}_{+\alpha_1\alpha_2}\over{2}}T_+^{\alpha_1\alpha_2} +
{{\tilde X}_{-\alpha_1\alpha_2}\over{2}}T_-^{\alpha_1\alpha_2}. }

Following the preceding subsection (and working with $e=0$ for
simplicity) we find that the charged states now have an extra $N_1N_2$
in their multiplicities, and have masses
\eqn\Nmasses{
\eqalign{
{\tilde X}^i_{\pm\alpha_1\alpha_2}:\quad m^2&={r^2\over 4\nu^2}\cr
{\tilde A}^9_{\pm\alpha_1\alpha_2}:\quad m^2&={r^2\over 4\nu^2}\cr
{\tilde A}^0_{\pm\alpha_1\alpha_2}+{\tilde X}
^1_{\pm\alpha_1\alpha_2}:\quad m^2&={r^2\over 4\nu^2} + {gp\over \nu}\cr
{\tilde A}^0_{\pm\alpha_1\alpha_2}-{\tilde X}
^1_{\pm\alpha_1\alpha_2}:\quad m^2&={r^2\over 4\nu^2} - {gp\over \nu}\cr
}}
\noindent where
\eqn\nudef{ {1\over \nu }= {1\over N_1} +{1\over N_2}\ .}
Likewise, the charged fermions and ghosts have masses as in
the $N=2$ case with the trivial rescalings to 
\eqn\pbreplace{
{\bar p} = {p\over 2\nu}\ ,\ 
 {\bar b} ={b \over 2\nu} }

Therefore, in the case of trivial boundary conditions the amplitude (and
hamiltonian) is
exactly as computed in \onephase\ with the only 
difference being multiplication by $N_1N_2$ and replacement of $p$ and
$b$ as in \pbreplace. Note that $2{\bar p} = {{p}\over{N_1}} +
{{p}\over{N_2}}$ is simply relative velocity of the two
clusters, and ${\bar b} = {{1}\over{2}} ({{b}\over{N_1}} + 
{{b}\over{N_2}})$ is precisely the impact parameter between the
clusters.

In the case of long-string boundary conditions, this result is modified.
Now
\eqn\lsbc{
X(\sigma+2\pi)=\left\lgroup\matrix{V^{-1}_1&0\cr
0&V^{-1}_2\cr}\right\rgroup
X(\sigma)\left\lgroup\matrix{V_1&0\cr
0&V_2\cr}\right\rgroup, }
and in particular the charged off-diagonal blocks satisfy
twisted boundary conditions
\eqn\TwBC{
\eqalign{
X_+ (\sigma + 1)&=V_1^{-1}X_+ (\sigma) V_2\cr
X_- (\sigma + 1)&=V_2^{-1}X_- (\sigma) V_1\ .\cr} }
\noindent The matrices $V$ can be diagonalized by working on basis
vectors
\eqn\Newbasis{
w_k={{1}\over{\sqrt{N}}}\left\lgroup\matrix
{1\cr
\lambda^k\cr
\lambda^{2k}\cr
\vdots\cr
{\lambda}^{(N-1)k}\cr}\right\rgroup\ ,\ \lambda=e^{2\pi i/N},\ k\in{\bbb Z}}
and in this basis simply give phases ${\lambda^k}$. Thus the
amplitude \onephase\  is modified to 
\eqn\Nonephase{\eqalign{
i\delta_1= \hf \sqrt{i\over 2\pi}
\sum^{N_i}_{\alpha_i=1}\sum_n\,\, 
e^{-in\left[2\pi(\alpha_1/N_1-\alpha_2/N_2)+a\right]}
\int^\infty_0&\,\,{{dT}\over{T}}
{e^{in^2/4T-ig^2{\bar b}^2T}}{1\over \sqrt{T} \sin(g{\bar p} T)}\cr
&\biggl\lbrack -6-2\,{\cos} (2g{\bar p}T) + 8\,\cos(g{\bar
p}T)\biggr\rbrack\ ,}}
and the interaction hamiltonian takes the form
\eqn\intham{H_{int}\approx {g^4p^4\over 2\pi i}
\sum_{n,\alpha_1,\alpha_2} 
e^{-in\left[2\pi(\alpha_1/N_1-\alpha_2/N_2)+a\right]}
\int {dT} T^2
e^{in^2/4T - ir^2T}\ .}

For non-zero $n$, the supergravity 
correspondence no longer holds when $ N>2 $: the matrix string 
then yields a different result.
%due to the fact that we have to sum over 
%the fractional Fourier frequencies $n_1$ and $n_2$, which has the
%effect of restricting the minimal exchanged D-particle number
%to $N_1N_2$ (in case $N_1$ and $N_2$ are relatively prime). 
In \intham\ the expression in the summation only gives a non-zero 
contribution  when the integer $n$ is a multiple of both $N_1$ and $N_2$. 
Hence we see that the minimal exchanged D-particle number 
between two long strings of length $N_1$ and $N_2$ must be proportional to
$N_1N_2$ (if the lengths are relatively prime), else the amplitude will 
be simply zero. 

This leads us to the conclusion that the long strings do not give an effective
means of creating D-particles. For two strings to create a minimally charge
D-pair, the strings apparently first need to each emit a minimal length string
%fractionate into a collection 
%of $N_1$ respectively $N_2$ short strings, 
such that two short strings of 
both collections can exchange a single D-particle. In the SYM language,
this last process is effectively an $SU(2)$ process, where correspondence
with 11-dimensional supergravity is found. It is important to note, however,
that the electric flux thus created must then subsequently spread out
over the complete $U(N_i)$ gauge group, since otherwise it would not carry
the SYM energy appropriate for the massive D-particle with $p^+ = N_i/R$.

Furthermore, in the sector with a fixed $p^+$ momentum, we now have an
improved
idea of what state contributes most to D-charge production: it is the state
with trivial boundary conditions, \trivbc, corresponding to a collection of
minimal length strings.  Since this is the state that yields amplitudes
agreeing with low-energy supergravity in the limit $g\rightarrow 0$, it is
apparently this state (or a bound version of it when finite $g$ effects are
taken into account) that dominates the wavefunction of the graviton in the
small $g$ region, rather than the state with the long
string boundary conditions \stringbc.  

\newsec{Ranges of validity}

In this section we will give a preliminary discussion of the relevant
scales and ranges of validity of the calculations of the preceding
sections.  This analysis is preliminary in that the systematics of the
perturbation theory for the Yang-Mills lagrangian \finact\ has not
been performed at the level of that for pure Matrix theory\refs{\BBPT} and
additional subtleties are possible.  We leave such analysis for future work.
For simplicity we will consider the case where the $p^+$ momentum of the
two incoming states are comparable, $N_1\sim N_2\sim N$.  Our arguments
readily generalize to the probe situation $N_1\gg N_2$

We begin by considering the expansion of the action about a classical
background as in \sdecomp; such a treatment is relevant both for corrections
to the saddlepoint solutions of section four as well as for the systematic
treatment of D-pair production.  

This expansion is governed by the Yang-Mills coupling $\gym$, and naively one
expects the condition $\gym\ll1$ for corrections to be small.  However, as
mentioned in the introduction, the Yang-Mills coupling is scale dependent
and one expects the relevant scale to be set by the physics one is
considering.  

For example, in the scattering with background \initstate, loops of the
charged, 
massive states of the YM theory play a central role.  One either has a
loop localized on the cylinder, whose calculation leads to the ${\cal
O}(v^4/r^6)$ supergravity potential, or the loop can encircle the cylinder
leading to the D-pair production that we have computed.  These massive states
receive masses of minimum size $b/g_s$ through the Higgs effect, setting the
length scale $\ell_b\simeq g_s/b$.  At this scale, we expect the relevant
dimensionless parameter to be 
\eqn\gymb{ \gym \ell_b \simeq {1\over b }\ .}
Smallness of this parameter thus requires 
\eqn\bcondit{b\gg1\ .}

For the case of pair creation, there is another requirement arising from
the condition that the backreaction due to the created electric field be
small.  One way of stating this is to require that the YM energy be large
as compared to the energy stored in the electric field,
\eqn\YMenerg{ {p^2}\gg \gym^2\ .}

Finally, in the case of the string interactions of section four, we see from
\rlimitp\ that the relevant scale is set by the parameter $|B|$, and is given
by 
\eqn\Bscale{ \ell_{inst} \simeq \left({g_s\over |B|}\right)^{2/3} = \left(
{g_s^2 N\over p^2 \sin \theta}\right)^{1/3} 
 \ .}
At this scale the dimensionless coupling is given by
\eqn\Bymc{ \gym \ell_{inst} \simeq \left({N\over g_s p^2 \sin
\theta}\right)^{1/3}\ .}
Another condition to apply the methods of section four is that the size of
the instanton be small as compared to the size of the cylinder,
$\ell_{inst}\ll 1$, or 
\eqn\smallinst{ p^2 \gg {g_s^2 N\over \sin\theta}\
.}

It is certainly possible to simultaneously satisfy the conditions \bcondit,
\YMenerg, and \smallinst, as well as the more stringent condition $\gym\ll1$, 
for finite $N$ and large $s\sim p^2$.  If all important corrections are
governed by expansions in the parameters of \gymb\ and \Bymc, then it
appears possible to even push the calculations into the range
$g_s\roughly<1$.  

A more complete analysis can be performed in the large $g_s$ (large $R_9$)
case 
in the restricted energy range
\eqn\Matrange{ {1\over g_s} \ll  E \ll g_s \ .}
The
lower bound corresponds to the energy threshold to create D charge, and the
upper bound is the energy to create winding states wrapping $x^9$.  In
between these bounds the theory can be effectively described by Matrix
theory DLCQ quantized in 10 dimensions.  

As explained in \refs{\BBPT}, the Matrix expansion is an expansion in 
terms of the form
\eqn\matexpp{\left({N\over r^3}\right)^L \left(v^2\over R^2 r^4 \right)^n}
where $L$ counts loops.  The terms with $L=n$ are readily identified with
terms in the corresponding supergravity expansion, and the small parameter
justifying this expansion is\refs{\BBPT,\BGL}
\eqn\sugrapar{ {N v^2\over R^2 r^7}\ll1\ .}
This has a simple physical interpretation, which is easily seen by
estimating the net transverse momentum transfer due to the potential
\eqn\approxpot{{N^2 v^2\over R^3 r^7}\ ;}
this gives 
\eqn\transvp{ \Delta p_\perp = {N^2 v^3\over R^3 b^7}\ .}
The condition \sugrapar\ is then easily seen to be $\Delta p_\perp \ll p$,
or equivalently $\theta\ll 1$ where $\theta$ is the scattering angle.
Combining this with the threshold condition \Matrange\ then yields
\eqn\gpsll{g_s p \gg 1\ , }
in correspondence with \Matrange.

Expansion terms with $n>L$ are then suppressed for 
\eqn\nexpp{ p^2 \ll N^2 r^4 \ ,} 
and terms with $n<L$ for 
\eqn\Lexpp{r \gg (Ng_s)^{1/3} \ .}
It is unclear whether the latter condition is strictly necessary; the first
term in this expansion vanishes \refs{\BeBe,\BBPT}, and the other terms
have been conjectured to vanish in \refs{\BGL}.

To better understand these conditions, we convert them into statements
relating the Mandelstam parameter $s\sim p^2$ and the angle $\theta$.  It
is easily seen that condition 
\nexpp\ becomes 
\eqn\ntheta{s \ll N^2 \left({N\over \theta}\right)^{4/3}\ }
and the condition \Lexpp\ becomes 
\eqn\Ltheta{s\gg g_s^{7/3} N^{10/3} \theta\ .}

Comparing \smallinst\ with \ntheta\ and \Ltheta, we see that within the
energy range given by \Matrange\ the instanton and D-particle production
calculations are not obviously simultaneously valid.  However, outside of
this range, we appeal to the preceding (less rigorous) analysis which
suggests that these calculations are indeed simultaneously valid at large
$s$, and may even be extendable to $g_s\roughly<1$.  It is partly on this
basis that we will, in the next section, consider the consequences of
combining these two calculations.

\newsec{Discussion and Conclusions}

We begin this section by recalling several observations from our preceding
discussion.  The first is that, as pointed out in section 6.4, string
scattering only efficiently produces D charge if the strings break off at
least one minimal length string.  Furthermore, sections four and five
discussed saddlepoint configurations that are expected to make important
contributions to high energy, fixed angle scattering.  Combining these
yields a picture of how the important non-perturbative process of D-charge
production can arise in high-energy string scattering.

The analysis of Gross and Mende\refs{\GrMe} found saddlepoints believed to
dominate scattering at high energy.  These saddlepoints have a common
structure at arbitrary genus, and the contributions of these
saddlepoints grows with the genus suggesting the relevance of
non-perturbative effects.  We have found a new version of their analysis in
which a mechanism appears that can cut off this growth.  The cutoff
originates from the minimal string length, which is in our language the
minimal $p^+$.  String fragmentation is 
stopped when the string breaks into the maximal
number of minimal-length strings.\foot{In this sense, the matrix
string formalism behaves very similar to
the discretized models of string theory,
advocated in particular by C. Thorn. We
thank S. Shenker  for drawing our
attention to this similarity.}

It is precisely in the context of minimal length string scattering that we
have found that D-charge pair production can become an important effect.
We therefore have a very nice picture in which the instantons of section
four and five lead to maximal fragmentation of the strings, and this is 
followed by
the production of D-charge via the process of section six.  
Here we expect that the size \rlimitp\ of the instanton,
as well as the corresponding minimal distance \dlimit,
may be an important ingredient in determining the size of
both these effects.

From the
stringy viewpoint this is an intrinsically non-perturbative process.  This
is suggestive that there is in fact a basic connection between these two
processes, and in particular that the non-perturbative production of
D-charge is an important correction to the high-energy scattering analysis
of \GrMe.  While we believe that, by combining the various ingredients 
presented in this paper, it may be possible to obtain definite quantitative
estimates of these corrections, we leave further analysis of this connection 
for future work.

Next we turn to several other observations and connections.

First, recall that Banks and Susskind \refs{\BaSu} previously considered
the $D{\bar D}$ system in the context of perturbative string theory.  There
they found an instability with unknown outcome.
In the present framework we have been able to treat the same system
analytically, at least in the large $g_s$ limit, without signs of
pathology.  In principle, the Matrix string calculations appear to extend
to arbitrary $g_s$.  
One might hope that some extrapolation of our approach could shed
further light on the discussion of \BaSu.

It is an interesting conceptual question under which circumstances
one needs to include the virtual effects of D-particles propagating in loops.  
Although in the literal sense of an expansion about $g_s=0$ they do not
contribute, since they have infinite mass there, there is clearly a 
strong sense in which D-particles can be found in intermediate states 
when $g_s$ is finite.  
Indeed, intermediate states with D-charge are distinguished
from other intermediate states only by the presence of electric flux, and
there is no apparent reason why these should be suppressed at finite $g_s$.
In fact, looking at the results of section 6, leads one to suspect that 
it may be possible to extend the matrix string interactions as summarized 
in section 2.3 to include the possibility of electric flux ``pair'' creation. 
The eleven dimensional symmetry of M-theory, in particular,  
suggests as a possible generalization of the DVV string-interaction vertex,
an expression of the form $V_{int} = V_{twist} \delta(A_{12})$ 
(with $A_{12}$ the difference between the U(1) gauge fields on the two 
strings that are created). With this choice of vertex, the couplings
between string and all $n$-D-particle bound states are all of the same 
strength. This would suggest that there may possibly exist a systematic
semi-classical expansion in string theory -- generalizing the standard
perturbative expansion -- in which the D-particle-loop contributions play
the role of instanton-like corrections. Indeed, in other recent 
studies of non-perturbative contributions to string scattering 
amplitudes \refs{\Greo\GrVh\GGV-\Gret} it was suggested that D-particle loops 
are related via T-duality to D-instanton contributions in IIB string 
theory. It clearly would be interesting to see if the suggestive
formulas obtained in these works can possibly be reproduced via the 
Matrix string methods developed in this paper.

To conclude, we have succeeded in using the Matrix String approach to
begin an investigation of 
aspects of high energy string scattering, and in particular to
begin to explore the role of important non-perturbative (from the string
viewpoint) processes such as D-charge production.  Further investigation
along these lines is expected to unravel a rich structure at substringy
scales, and may shed further light on the short distance structure and
fundamental degrees of freedom and dynamics of M-theory.

\bigskip\bigskip\centerline{{\bf Acknowledgements}}\nobreak
This work was supported in part by DOE grant DE-FG03-91ER40618,
by NSF PYI grant PHY-9157463, by the Dutch Royal Academy of Sciences,
a Pionier Fellowship of NWO, an A.P. Sloan Fellowship, and by the Packard
Foundation.  We wish to thank S. Bais, U. Danielsson, 
R. Dijkgraaf, M. Douglas, A. Hashimoto, I. Kogan, G. Lifschytz, H. Nicolai, 
A. Nudelman, H. Ooguri, J. Polchinski, P. Pouliot, L. Randall, B. Schroers,
J. Striet, A. Strominger, E. Verlinde, E. Witten, T. Wynter, and
especially D. Gross and S. Shenker
for very valuable conversations.  
This work was initiated 
at the Aspen Center for Physics, whose hospitality we gratefully
acknowledge.  H.V. also wishes to thank the Institute for Theoretical
Physics at Santa Barbara for its hospitality.

While we were writing up our results, we received \refs{\KVK}, which has
some overlap with the discussion of section six.

\appendix{A}{The Matrix String Approach at Finite N.}

Most of our calculations in the main text are performed 
for type IIA string theory in the DLCQ approach, in
other words for the IIA string compactified on a lightlike circle of
``radius'' $R$
\eqn\dlcq{
x^- \equiv x^- + 4\pi R}
This means that from the M-theory point of view we have
compactified on $S^1\times S^1$, with the second compactification on a
circle whose radius $R_9$ is expressed in terms of the corresponding 
string length as $R_9 = g_s \ell_s$. Through most of the paper,
we work in string units $\ell_s=1$, so that $R_9=g_s$.    

In this Appendix we will rederive the 1+1 dimensional $U(N)$ SYM
lagrangian as the appropriate Matrix formulation of M-theory in this
discrete light-cone limit.  This is useful in understanding the precise
scalings and relations between parameters.  
We will follow closely the reasoning of 
\refs{\Seib,\Sen,\Tayl}.

To begin with, consider \mth\ compactified on a circle with radius
$\Rele\ll\ltp$, where $\ltp$ is the eleven-dimensional Planck scale.
Suppose we restrict attention to the sector with momentum $\pele=N/\Rele$.
The lowest excitations in this sector occur at energies $E\sim \Rele
\Mtp^2$, and are described by the action (here for simplicity we omit
fermions)\refs{\DFS,\KaPo,\DKPS}
\eqn\matact{S=\int {dt\over 2\Rele} {\rm Tr}\left({\dot X}^2 + \hf\Rele^2
\Mtp^6[X,X]^2\right)\ .}
$X$ is an $N\times N$ matrix; the diagonal terms in $X$ correspond to the
coordinates of the branes, with kinetic energies
$M_{D0}\Xdot^2/2=\Xdot^2/2\Rele$ and the off-diagonal components describe
the creation of straight strings stretched between the branes, with
characteristic energies $\sim |x_i-x_j| \Rele \Mtp^3$.  This action
neglects string oscillations and higher energy excitations (Planck scale
excitations, brane creation).  The lowest of these begin at the string
scale (defined with respect to the compactification on $\Rele$)
\eqn\Mtsdef{\Mts= \sqrt{\Rele \Mtp^3}\ .}

Next we compactify on a circle of radius $\Rtn$, through the
identification $X_9\simeq X_9 + 2\pi \Rtn$.  In the matrix context, Taylor
\refs{\Tayl} has argued that this is most easily described by passing to
the dual circle of radius $R_9' = 1/2\pi\Rtn \Mts^2$.  The $X_9$ variables
dualize to a gauge field through the identification
\eqn\Adef{A_x=\Mts^2\sum_n e^{inx/R_9'}X^9_{0n}}
and the wrapped strings combine into a new definition of the
remaining fields,
\eqn\Ydef{\Xt^i = \sum_n e^{inx/R_9'} X^i_{0n}\ ,}
with the resulting action 
\eqn\dualact{S= \int {dt\over 2\Rele} \int_0^{2\pi R_9'} {dx\over 2\pi R_9'}
{\rm Tr} \left\{\Xtdot^2 + {1\over \Mts^4} \Adotx^2 - \left(\partial_x \Xt +i 
[A_x,\Xt]\right)^2 +\hf \Mts^4[\Xt,\Xt]^2\right\}\ .}
It's then convenient to redefine the spatial coordinate,
\eqn\sigdef{x=2\pi R_9' \sigma\ ;\ A_x = {A_\sigma\over 2\pi R_9'}\ , }
giving
\eqn\newS{ 
S= \int {dt\over 2\Rele} \int_0^{1} {d\sigma} {\rm Tr} \left\{ \Xtdot^2
+ \Rtn^2 \Adot_\sigma^2 - {\Rtn^2\Rele^2\over \ltp^6} (D_\sigma \Xt)^2 +
\hf{\Rele^2\over  \ltp^6} [\Xt,\Xt]^2 \right\} \ .}

The action \newS\ so far corresponds to the low-energy action of the system
of N D0 branes on the circle of radius $\Rele$.  The next step is to take
the limit 
\eqn\sclim{\Rele\rightarrow0, \Mtp\rightarrow\infty}
holding
\eqn\scaling{\Mtp^2R_{11} \equiv \Mpl^2R}
fixed, following \refs{\Seib,\Sen}.  This gives the infinite momentum
limit, $p_{11}\rightarrow\infty$.  At the same time, one
defines a rescaled $X$ and $R_9$,
\eqn\rescale{X={\Mtp\over \Mpl} \Xt\ ;\ R_9 = {\Mtp\over \Mpl} 
{\tilde R_9}\ .}
In the limit \sclim,\scaling\ all the higher
excitations mentioned above (string, Planck, and $D0$) decouple and the
lagrangian \newS\ becomes exact.  Written in terms of the rescaled
variables,
\eqn\Sdef{S=\int {dt\over 2R}\int_0^{1} {d\sigma}
 {\rm Tr}\left\{ \Xdot^2 +
R_9^2 \Asd^2 - {R_9^2 R^2\over \lpl^6} (D_\sigma X)^2 +\hf{R^2\over \lpl^6}
[X,X]^2\right\}\ .}

Eq.~\Sdef\ therefore describes DLCQ IIA string theory lightlike
compactified on a circle of ``radius'' $R$.  The physical string coupling
and string scale are defined by the usual relations 
\eqn\coupdef{ \Rn = g_s \lst\ ,\ \lpl= g_s^{1/3} \lst\ .}
Finally, as a matter of convenience it is easier to work in terms of a
rescaled time 
\eqn\taudef{\tau=t R/\lst^2\ ,}
and to redefine 
\eqn\xrescale{ X\rightarrow \lst X\ ,}
corresponding to measuring $X$ in string units.  With these redefinitions
the action indeed takes the form as given in the introduction.  In the main
text we will take $\lst=1$.

In order to describe physical scattering processes, we will also need to
parameterize the external states.  We will separate off the two compact
momenta from the transverse momentum $\vecp$, and write 
\eqn\momdef{p^\mu=(p^-,\vecp,p_9=n/R_9,p^+=N/R)\ .}
For example, when we compute scattering of two gravitons into two $D0$
particles (now defined with respect to the $R_9$ compactification),
initially $n=0$ and finally $n=\pm1$.  Notice that scaling the momentum
according to \rescale, finite $p$ corresponds to infinite ${\tilde p}= \Mtp
p/\Mpl$.  However, the corresponding energy scale in the original
variables of \newS\ is $E\sim {\tilde p}^2 \Rele = R^2 p^2$ which remains finite as
the other excitations decouple in the limit $\Rele\rightarrow0$.
In the case of two particles in the center of mass frame each with one unit
of DLCQ momentum, and with relative momentum $\vecp$,
the asymptotic state is described by
\eqn\astate{ X= \hf\left(\matrix{R\vecp t+ \vecb&0\cr0&-R\vecp t
-\vecb}\right)\ .}
In  the rescaled units defined by \taudef, \xrescale, and $\lst=1$,
this becomes 
\eqn\extstate{X= \hf\left(\matrix{\vecp \tau+ \vecb&0\cr0&-\vecp \tau
-\vecb}\right)\ ,}
with the understanding that the external momentum and impact parameter are
measured in string units.

\listrefs
\end